\documentclass[%
 reprint,
superscriptaddress,https://www.overleaf.com/project/64994f20bd1198e1d7691150
 amsmath,amssymb,
 aps,pre,hyphens,floatfix
]{revtex4-1}

\usepackage{lineno}

\usepackage{graphicx}
\usepackage{subfigure}

\usepackage{float} 
\usepackage{parskip}
\usepackage{ulem}
\usepackage{hyperref}
\hypersetup{hypertex=true,
colorlinks=true,
linkcolor=blue,
anchorcolor=blue,
citecolor=blue}
\usepackage{enumerate}
\usepackage[usenames,dvipsnames,svgnames,table]{xcolor}
\usepackage{IEEEtrantools}
\usepackage{graphicx}
\usepackage{array}
\usepackage{multirow}
\usepackage{verbatim}
\usepackage{float}
\usepackage{amsmath}

\begin{document}

\preprint{APS/123-QED}

\title{
Critical dynamics of the directed percolation with L\'evy-driven temporally quenched disorder
}

\author{Yanyang Wang}
\affiliation{Key Laboratory of Quark and Lepton Physics (MOE) and Institute of Particle Physics, Central China Normal University, Wuhan 430079, China}

\author{Yuxiang Yang}
\email[]{yuxyang@mails.ccnu.edu.cn}
\affiliation{Key Laboratory of Quark and Lepton Physics (MOE) and Institute of Particle Physics, Central China Normal University, Wuhan 430079, China}
\author{Wei Li}
\affiliation{Key Laboratory of Quark and Lepton Physics (MOE) and Institute of Particle Physics, Central China Normal University, Wuhan 430079, China}
\affiliation{ESIEA, Campus Ivry sur Seine, 73 bis Avenue Maurice Thorez, 94200
Ivry sur Seine.}

\begin{abstract}

Quenched disorder in absorbing phase transitions can disrupt the structure and symmetry of reaction-diffusion processes, offering a more accurate mapping to real physical systems. We developed a temporally quenched disorder method in the (1+1)-dimensional direct percolation (DP) model, where the increment of conditional probability is determined by the cumulative distribution function (CDF) of the L\'evy distribution. Monte Carlo (MC) simulations reveal that the model has a critical region governing the transition between absorbing and active states, and this region changes as the parameter $\beta$, which influences distribution properties. Guided by dynamic scaling laws, we observe that significant variations in the L\'evy distribution parameter $\beta$ lead to notable changes in the particle density decay exponent $\alpha$, total particle number exponent $\theta$, and spreading exponent $\tilde{z}$. The quenching mechanism we introduced has broad potential applications in various theoretical and experimental studies of absorbing phase transitions.

\end{abstract}
\maketitle



\section{Introduction}
\label{intro}

Classical phase transitions involve both equilibrium and non-equilibrium systems, which are closely related both theoretically and experimentally. Even equilibrium systems in steady states may deviate from equilibrium or undergo relaxation processes due to external disturbances or shocks\cite{tauber2014critical}. Several theoretical methods, including mean-field theory\cite{cafiero1998disordered,1989Quantum}, density matrix renormalization, low-density series expansions, and low-order field theory approximations, applied to the Ising model, have also been used to study non-equilibrium systems\cite{christensen2005complexity,henkel2008non,tauber2014critical,2005field}. Moreover, Monte Carlo (MC) simulations have been adapted for non-equilibrium reaction-diffusion systems and have evolved into widely established simulation techniques\cite{lubeck2006crossover,zhong1995universality,ma2024emergent,deng2024chimera,dong2021optimal,liu2021efficient,qing2024time}.

The directed percolation (DP) model, with its robust stability, introduced the concept of universality classes\cite{grassberger1979reggeon,janssen1981nonequilibrium,1982On,vojta2005critical}. Representative models include, but are not limited to, interface growth\cite{tang1992pinning}, contact processes\cite{harris1974contact}, and turbulence\cite{pomeau1986front}. The evolution of reaction-diffusion systems, when determining the DP universality class, typically relies on fixed conditional probabilities that maintain the spatial and temporal stability of conservation laws\cite{hinrichsen2000non}. However, real-world physical systems often deviate from the DP model\cite{hinrichsen1999flowing,hinrichsen2000flowing,takeuchi2007directed,vojta2004broadening}. Analysis of correlation scales indicates that temporally quenched disorder may undermine the robustness of the DP universality class, thereby altering its critical properties \cite{hooyberghs2004absorbing,gonzaga2019quenched,dickman2009contact,fallert2009scaling,neugebauer2006contact,2005Low}.

\begin{figure*}[t]
    \centering
        \includegraphics[width=0.65\textwidth]{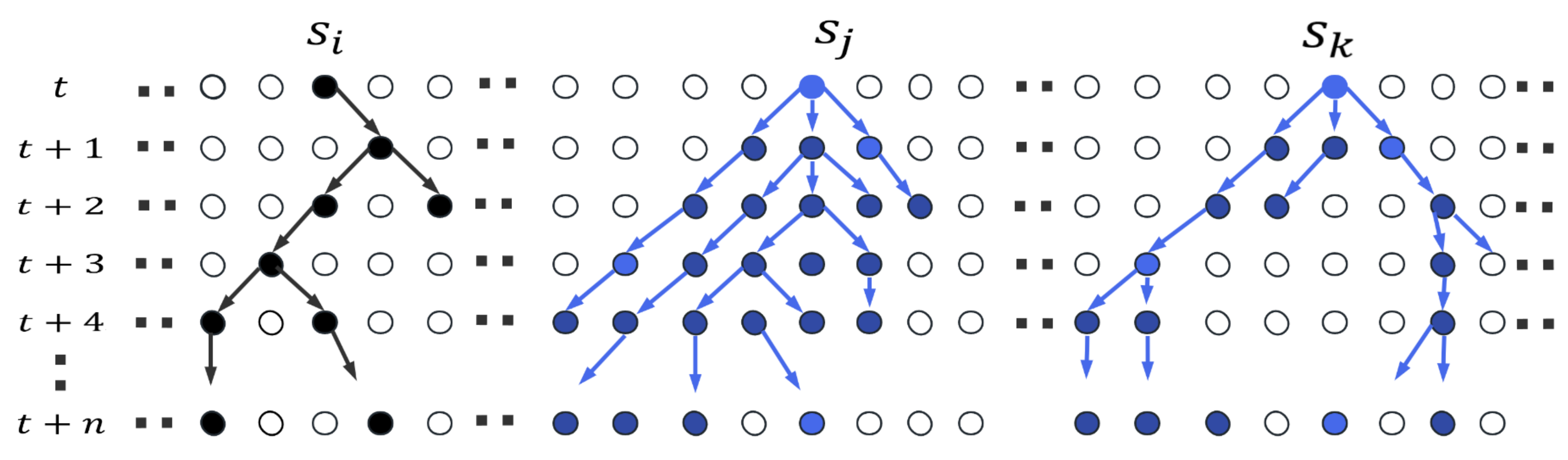} 
\caption{The evolution of the temporally quenched disorder DP system, compared to the standard DP process, is illustrated with examples. Here, $S_i$ represents the evolution of the standard DP process, $S_j$ represents the long-duration, high-percolation local multi-particle structures that may arise under temporally quenched disorder, and $S_k$ represents another possible effect, where large vacancy gaps appear under low conditional probabilities.}

\label{f1}
\end{figure*}

Introducing different forms of quenched disorder may cause changes in the system’s critical exponents, thereby enabling the exploration of the underlying variation laws. Most forms of quenched disorder in the above literature are relatively simple, with the noise increment of the conditional probability being randomly chosen at fixed intervals. Considering the disorder in real biophysical systems, introducing a quenched form with non-uniform random updates of noise increments may more accurately characterize the system’s evolutionary process on the time scale.

Lévy stable distributions, characterized by heavy tails, exhibit a higher propensity for extreme values than Gaussian distributions. This fat-tailed nature makes them ideal for modeling phenomena involving long-range interactions, nonlinear dynamics, and extreme fluctuations in stochastic processes. As the product of a generalized central limit theorem, they describe the limiting sum of independent and identically distributed random variables with diverging variance. Crucially, their tunable parameters allow a smooth transition between Gaussian and Cauchy distributions \cite{mantegna1994fast}, enriching physical interpretation and enabling exploration of diverse universality classes. Consequently, Lévy distributions find broad applications across various fields, including biological systems \cite{whitney2025generalapproachstatisticsmicrobial}, high-energy astrophysics \cite{aerdker2025superdiffusion}, and economics \cite{huang2024levyscorefunctionscorebased}.

Lévy distributions display pronounced power-law heavy-tailed features \cite{1998A,hinrichsen2007non}, inducing intermittent, intense perturbations in reaction-diffusion systems with quenched disorder. In epidemiology, sudden viral mutations or transmission surges from mass gatherings map to Lévy-driven quenched disorder fields. Simulations of these non-Gaussian quenched DP systems uncover nonlinear disease transmission dynamics, informing robust intervention strategies. In ecology, explosive species growth or mass extinctions triggered by abrupt environmental shifts are modeled by Lévy quenched disorder, yielding physical insights into ecosystem vulnerability and conservation. We compute probability increments using the cumulative distribution function (CDF) of the symmetric stable Lévy distribution and incorporate them into the conditional probability of the (1+1)-dimensional DP system to examine shifts in the critical point and exponents.

The structure of our paper is as follows: In Sec.\uppercase\expandafter{\romannumeral2}, we introduce the standard DP model and the method for incorporating L\'evy-driven temporally quenched disorder. Sec.\uppercase\expandafter{\romannumeral3} provides a brief overview of the dynamic scaling laws of absorbing phase transitions. Our experimental results and analysis are presented in Sec.\uppercase\expandafter{\romannumeral4}, where Sec.\uppercase\expandafter{\romannumeral4}.A shows a preliminary analysis of the phase diagram of the model's cluster graph. Sec.\uppercase\expandafter{\romannumeral4}.B provides a detailed explanation of the critical point determination process. In Sec.\uppercase\expandafter{\romannumeral4}.C, we measure the critical exponents $\alpha,\theta$, and $\tilde{z}$. Sec.\uppercase\expandafter{\romannumeral4}.D provides a summary and analysis of the results. Finally, Sec.\uppercase\expandafter{\romannumeral5} concludes the paper with a summary of our findings.

\section{Model}

This section first introduces the lattice simulation framework for the (1+1)-dimensional DP model, followed by a detailed elaboration on the procedure for incorporating time-quenched disorder driven by Lévy flights into the model.

In non-equilibrium statistical mechanics, DP is a stochastic multi-particle process\cite{potts1952some}. Monte Carlo simulations of the (1+1)-dimensional DP model are performed by updating discrete lattice sites. Typically, the system state is represented by binary values, where 1 indicates particle survival (active state) and 0 indicates extinction (absorbing state). The simulation is initialized with a fixed number of active seeds and evolves over discrete time steps. Once the initial conditions are set, the system evolves according to specific stochastic dynamics. The occupation state of site $S_i$ at the next time step is determined by the following update rule\cite{henkel2008non,1984Equivalence,1985Phase},
\begin{equation}
S_{i, t+1}= \begin{cases}1 & \text { if } S_{i-1, t} \neq S_{i+1, t} \quad \text { and } x_i(t)<p ,\\ 1 & \text { if } S_{i-1, t}=S_{i+1, t}=1 \text { and } x_i(t)<p(2-p), \\ 0 & \text { otherwise }.\end{cases}
\label{1}
\end{equation}

In the DP model, $x_i(t)$ is a random number between $[0,1]$, and $p$ is the conditional probability that governs the system's evolution. The value $S_{i,t+1}$ at the next time step represents the occupation state of site $i$. The state update of a site depends on the connection states of its neighboring edges\cite{henkel2008non}. 

Fig.~\ref{f1} illustrates the evolution of a (1+1)-dimensional DP system initiating from a single seed. The spatiotemporal pattern of sites $S_i$ demonstrates the typical particle growth clusters generated by a standard DP process. Under a sufficiently large conditional probability $p$, particles can survive from the initial state up to time step $t$, indicating the occurrence of percolation.

Considering that external noise and variations in unstable fields may cause the interaction strength to vary dynamically over time. For time-dependent quenched disorder in the DP system, the conditional probability $p$ can be modified as follows:
\begin{equation}
p \rightarrow p_t=p+\delta(t).
\label{2}
\end{equation}
Here, $\delta(t)$ represents the conditional probability increment at each time step during the DP evolution process. When the conditional probability value at each time step in the dynamic evolution process is altered, the evolution of the DP cluster may exhibit the features shown by $S_j$ and $S_k$ in Fig.~\ref{f1}, namely the aggregation of activated sites under high conditional probability and the formation of vacancies under low conditional probability.

The specific process of simulating the DP model and introducing temporally quenched disorder during its evolution is as follows. First, a one-dimensional lattice is created, and at $t=0$, each site of the lattice is traversed according to the update rules given by equations \ref{1} and \ref{2}. The conditional probability $p_t$ consists of two components: the conditional probability $p$ preset before the dynamic evolution of the DP model and the noise increment $\delta(t)$. To ensure the randomness and non-uniformity of $\delta(t)$, we use the step-length generation method based on the symmetric stable Lévy distribution \cite{mantegna1994fast}. The algorithm process for generating random step lengths $r_t$ that satisfy this distribution is as follows:
\begin{equation}
 r_t=\frac{u}{|v|^{1/{\beta}}},
 \label{3}
\end{equation}
where $u,v$ follow normal distribution
\begin{equation}
 u{\sim}N(0,{\sigma}_u^2),{\quad}{\quad}v{\sim}N(0,{\sigma}_v^2).
 \label{4}
\end{equation}
Additionally,
\begin{equation}
\sigma_u=\left\{\frac{\Gamma(\beta+1) \sin (\pi \beta / 2)}{\Gamma[(\beta+1) / 2] \beta 2^{(\beta-1) / 2}}\right\}^{1 / \beta}, \quad \sigma_v=1.
\label{5}
\end{equation}

The random step length $r_t$ is determined by two random numbers drawn from a normal distribution, with its absolute value taken to ensure positivity. Here, $r_t$ represents a time-dependent variable for the random step length within the evolutionary process of the temporally quenched disorder DP model. The parameter $\beta$ defines the exponent of the Lévy distribution and controls the scaling properties of the random step generation process.

Subsequently, we employ the Fast Fourier Transform to compute the integration of the L\'evy distribution,
\begin{equation}
{\psi}_{\beta}(r_t)=\frac{1}{\pi} \int_0^{\infty} \exp \left(-q^\beta\right) \cos (qr_t) dq,
\label{6}
\end{equation}
and its corresponding CDF $F(r_t)=\int_0^{r{_t}} {\psi}_{\beta}(r_t)dr_t$. According to evolutionary rules \ref{1} and \ref{2}, we introduce $F(r_t)$ as the noise increment $\delta(t)$ into the DP model evolution process and repeat the above random procedure at each subsequent time step $t>0$. By obtaining a random conditional probability at each distinct time step during the DP evolution, we investigate the influence of temporally quenched disorder on the system’s dynamics.

\begin{figure*}[t]
\begin{tabular}{cc}
    \includegraphics[width=0.14\textwidth]{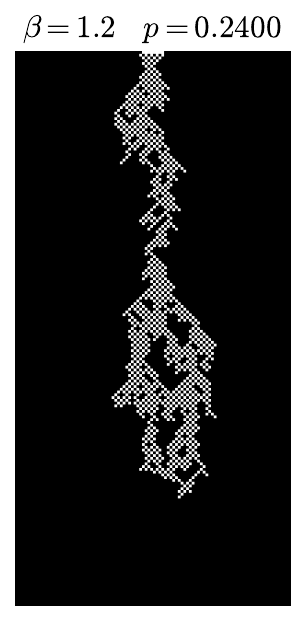} &
    $\qquad$\includegraphics[width=0.14\textwidth]{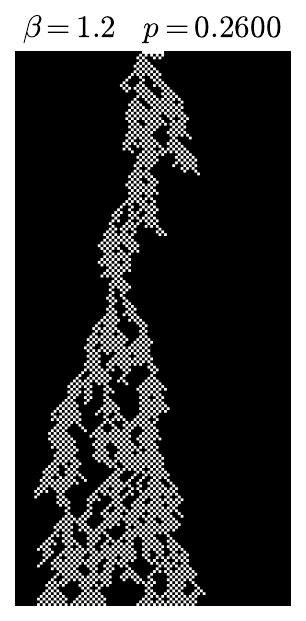} \\
    {\quad}{\quad}(a) & $\qquad$ {\quad}{\quad}(b)
\end{tabular}
\begin{tabular}{cc}
    \includegraphics[width=0.14\textwidth]{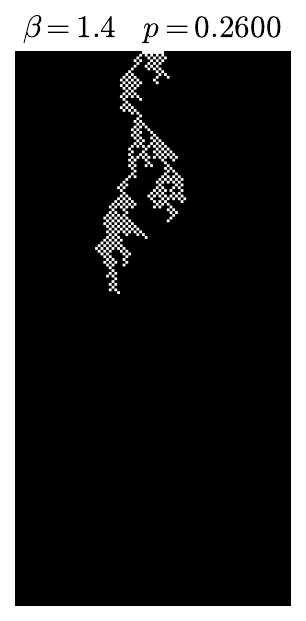} &
    $\qquad$\includegraphics[width=0.14\textwidth]{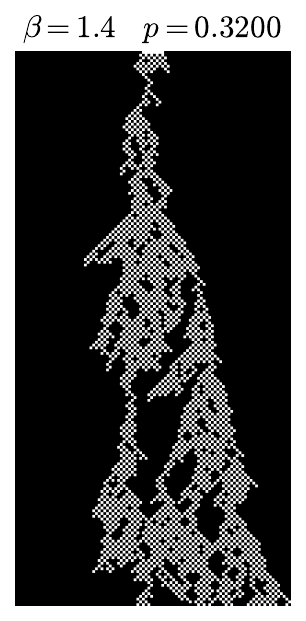} \\
    {\quad}{\quad}(c) & $\qquad$ {\quad}{\quad}(d)
\end{tabular}
\caption{Cluster diagrams of the L\'evy-driven temporally quenched disorder DP system are shown. (a) and (b) illustrate that, for the same $\beta$ value but different conditional probabilities $p$, the cluster diagrams exhibit a transition from the absorbing state to the active state, along with the appearance of large vacancy gaps and edge "avalanche" phenomena. (c) and (d) illustrates the phase transition behavior as the conditional probability changes, with different control parameters than those in (a) and (b). These phenomena indicate that the phase transition characteristics of the temporally quenched disorder DP system are influenced by the properties of the L\'evy distribution.
}
\label{f2}
\end{figure*}

\section{Method}

Absorbing phase transitions exhibit scaling properties that can be analyzed using scaling functions. Under the homogeneous initial condition, at the critical point, the time evolution of the particle density at active sites is described by
\begin{equation}
 {\rho(t)}{\sim}t^{-\alpha}.
\label{e_7}
\end{equation} 

Based on this dynamic scaling relation, for an absorbing phase transition system away from the critical point, the time evolution of particle density will deviate from a power-law decay. Therefore, this scaling law can be used as a criterion for determining the system's critical position, and under homogeneous initial condition, random errors are significantly reduced.

In numerical simulations, the critical dynamics of absorbing phase transitions are inevitably influenced by finite-size effects. Based on finite-size scaling theory, at the critical point, the particle density satisfies the following scaling relation:
\begin{equation}
\rho(L) \sim L^{-{\alpha}/z},
\label{e_13}
\end{equation}
where $z$ is the dynamical exponent, typically described in relation to system size as $L \sim \xi_{\perp} \sim t^{1/z}$, which characterizes the critical power-law expansion of the spatial correlation length under finite-size effects. After determining the critical point through numerous simulations of the particle density $\rho(t)$, which exhibits high sensitivity in the critical region, data collapse plots of the variables $\rho(t)t^{\alpha}$ and $t/L^{z}$ can be generated. This not only allows for the determination of the dynamical exponent $z$ but also serves as a verification for the accuracy of the critical point and the critical exponent $\alpha$.

In addition to simulating the reaction-diffusion process under the homogeneous initial condition, evolving clusters from locally active seeds can also simulate the absorbing phase transition and measure critical exponents\cite{henkel2008non}. When the system is at the critical point, the average number of active particles $N(t)$ and the mean squared spreading $R^2(t)$ follow: 
\begin{equation}
 { N(t)}{\sim}t^{\theta},{\quad}{\quad} R^2(t) {\sim}t^{\tilde{z}}.
\label{e_14}
\end{equation} 
$R^2(t)$ represents the mean square displacement of all active particles at time $t$ from the initial active seed. The critical exponents $\theta$ and $\tilde{z}$ are referred to as the initial slip exponent and the spreading exponent, respectively. The spreading exponent and the dynamical exponent $z$ satisfy the relation $\tilde{z} = 2/z$. Due to the high sensitivity of $N(t)$ near the critical point, it is commonly employed to locate the critical point and determine the critical exponent $\theta$. The dynamical exponent $z$ can typically be obtained indirectly under single-seed simulation conditions by measuring the power-law relationship of the mean-squared displacement.

When the system is not at the critical point, the average number of active particles in all clusters satisfies the scaling form
\begin{equation}
N(|p-p_c|) \sim  |p-p_c|^{\theta\nu_{\parallel}}.
\label{e_15}
\end{equation}
Data collapse plots of the variables $N(t)t^{-\theta}$ versus $t|p-p_c|^{\nu_{\parallel}}$ can be used to verify the accuracy of the critical point and the critical exponent $\theta$, and to determine the exponent $\nu_{\parallel}$ related to the temporal correlation length \cite{henkel2008non}.

\begin{figure*}[t]
\begin{tabular}{cc}
    \includegraphics[width=0.40\textwidth]{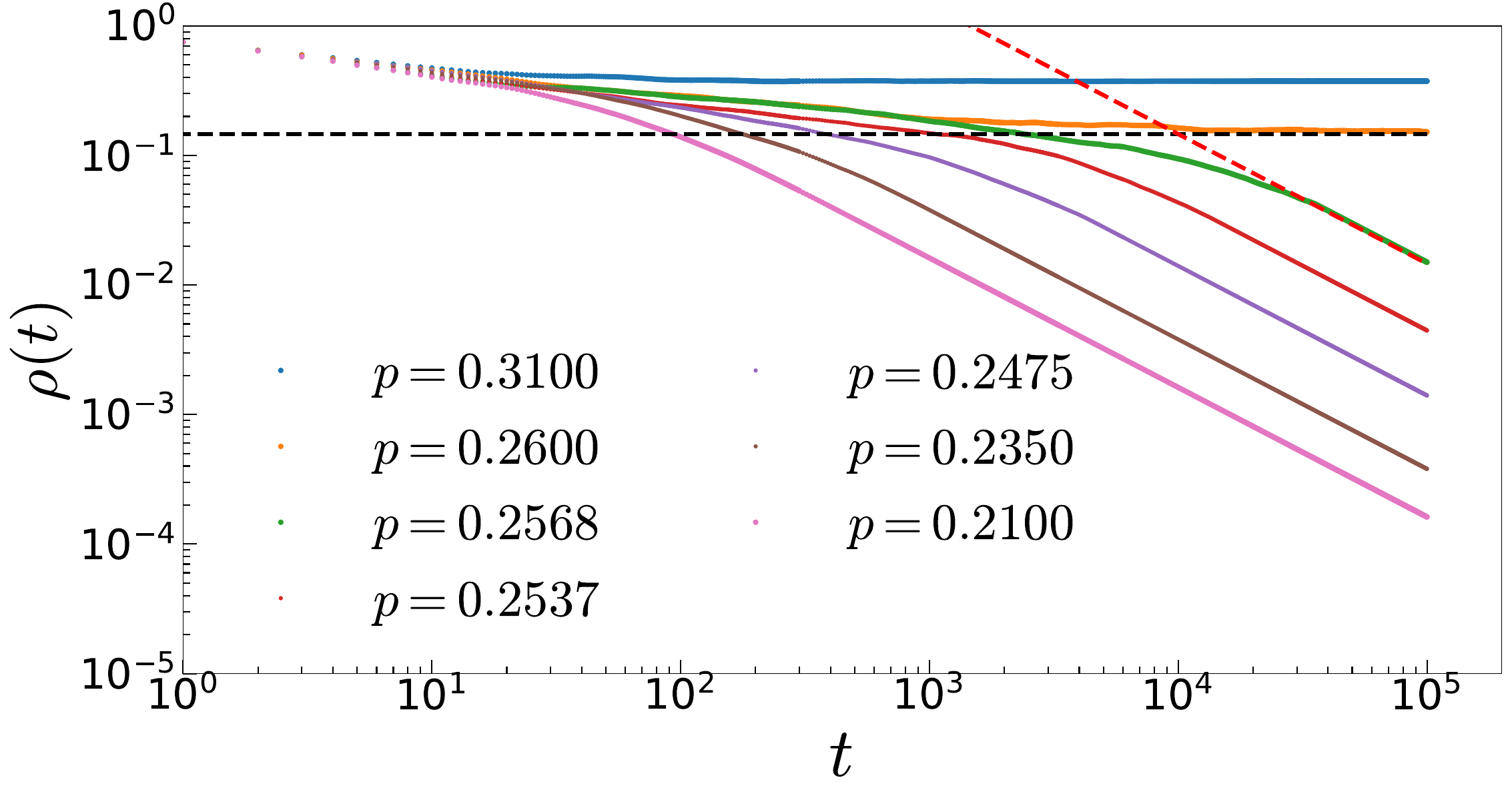} &
    $\qquad$\includegraphics[width=0.40\textwidth]{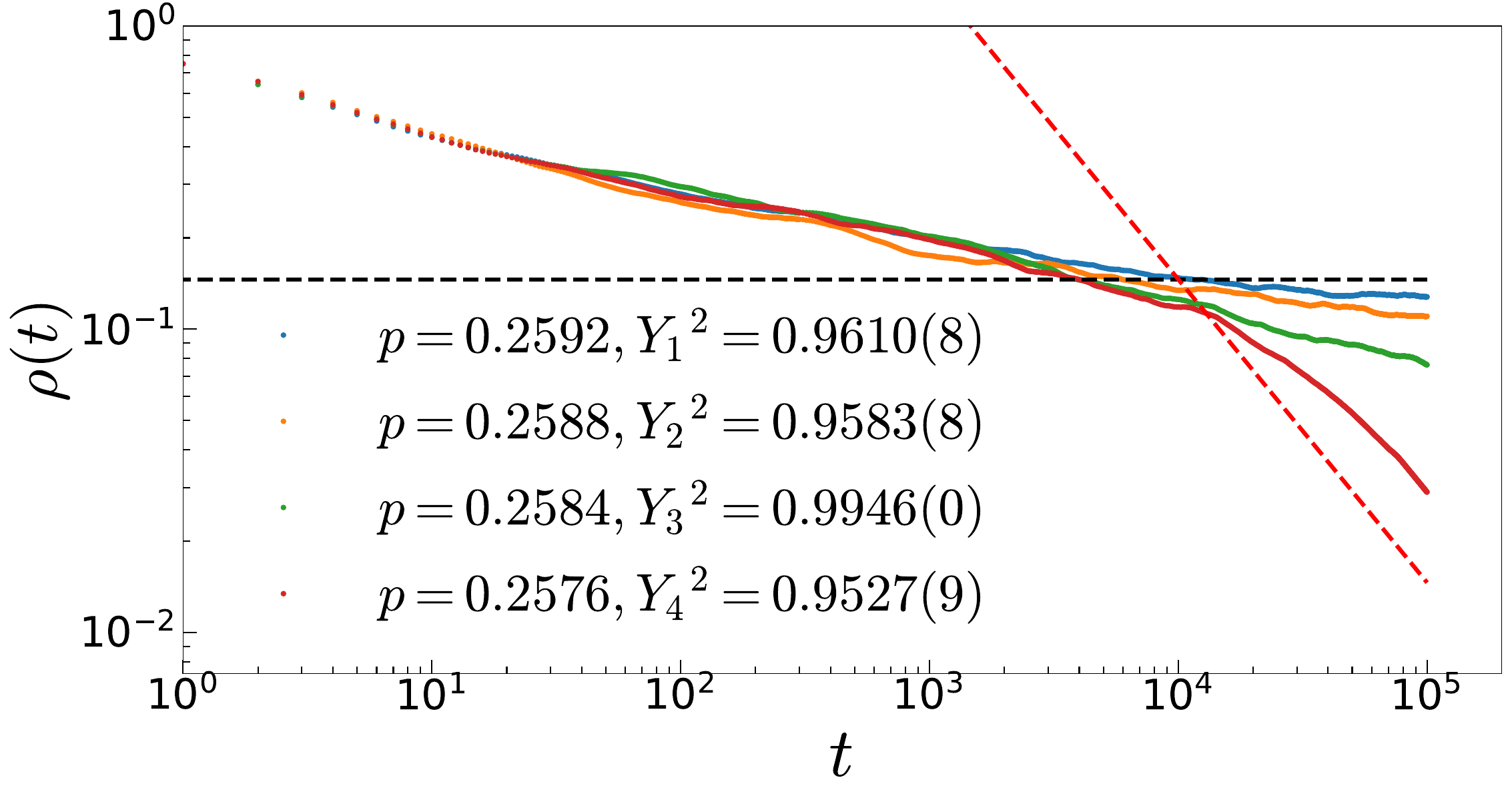} \\
    {\quad}{\quad}(a) & $\qquad$ {\quad}{\quad}(b)
\end{tabular}
\caption{(a) With reference to the dashed line, the critical region is gradually narrowed using the bisection method. Within the interval $[0.2568, 0.2600]$, the system’s average particle density still does not exhibit a global power-law behavior. (b) As the critical region is further narrowed, the critical point is determined using the goodness of fit. The value $p=0.2584$ corresponds to the optimal goodness of fit, making it a reliable reference for determining the critical point.}
\label{f_3}
\end{figure*}

\section{Result}
\subsection{Analysis of phase diagram features}

We first verify whether temporal quenched disorder with a specific distribution retains the general features of an absorption phase transition.

As shown in Fig.~\ref{f2}(a) and (b), we chose a L\'evy distribution with the control parameter $\beta =1.2$. When the probability is small, the system eventually reaches the absorbing state after sufficient time evolution. When the probability is large, the cluster graph shows the active state, with particles surviving for an extended period. This suggests that under temporally quenched disorder with a specific distribution, the DP system retains the characteristics of the absorbing phase transition.

When adjusting the quenching control parameter $\beta$, the system originally in the active state transitions to the absorbing state. As shown in Fig.~\ref{f2}(b)(c), when the conditional probability is set to $p = 0.2600$ and $\beta = 1.2$, the system remains in the active state, whereas at $\beta = 1.4$, the system enters the absorbing state. Moreover, when theconditional probability is set sufficiently high, such as $p = 0.3200$ in Fig.~\ref{f2}(d), the system again exhibits characteristics of the active state. 

We expect that the value of $\beta$ will strongly influence the spatiotemporal evolution behavior of the quenched DP system, as well as its critical point and critical exponents.

\begin{figure*}[t]
\begin{tabular}{cc}
    \includegraphics[width=0.40\textwidth]{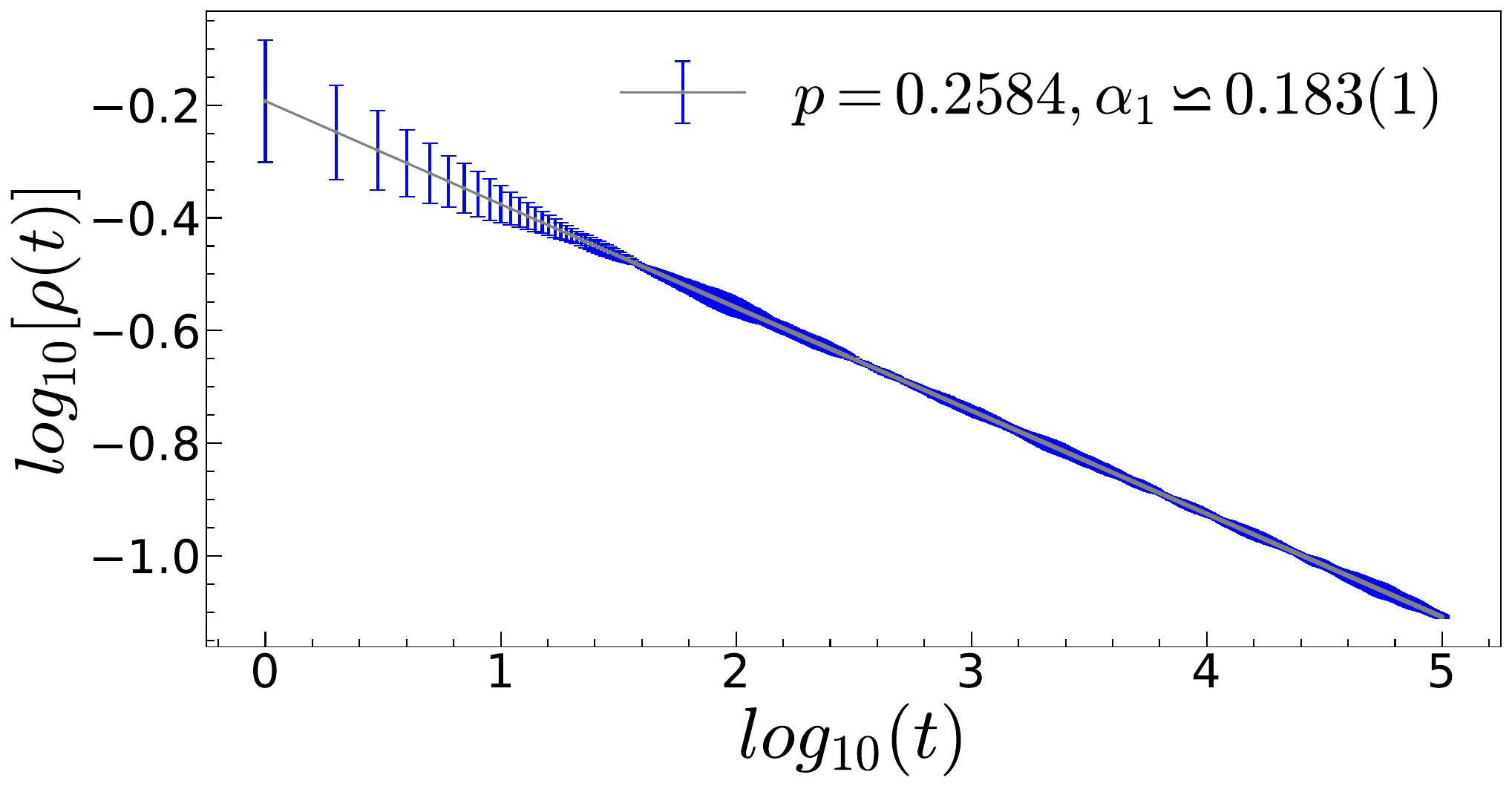} &
    $\qquad$\includegraphics[width=0.40\textwidth]{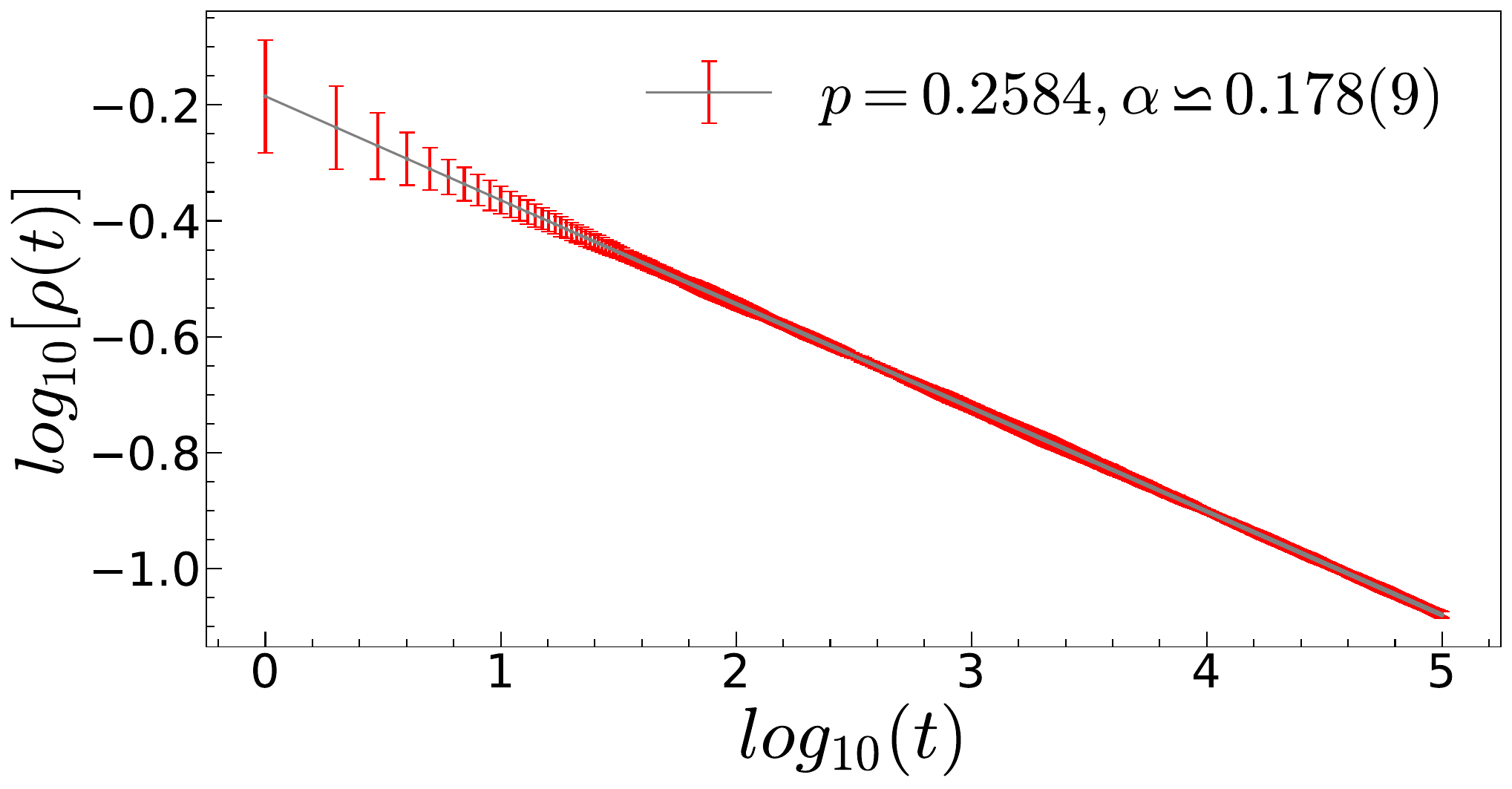} \\
    {\quad}{\quad}(a) & $\qquad$ {\quad}{\quad}(b)
\end{tabular}
\caption{Supplementary analysis of the power-law fitting for the temporal evolution of particle density at $\beta = 1.2$ and $p=0.2584$. (a) Results from a single simulation run comprising $200$ independent clusters on a $10^4 \times 10^5$ lattice. The error bars represent confidence intervals defined by the absolute values of the fitting residuals. (b) The average of five independent measurements. Compared to the single-run outcome in (a), the averaged confidence intervals exhibit greater stability and a significant narrowing. This improvement effectively enhances the reliability of the critical point determination.}
\label{f_4}
\end{figure*}

\subsection{Determination of the critical point}

In this section, we determine the critical point by narrowing the critical interval and using fitting validation, and present error analysis for confirmation.

Under the all-seeds condition, the periodic boundary conditions are equivalent to numerous repeated evolutions of single-seed systems to reduce statistical errors. We first conducted preliminary experiments using a system of size $L=10^4$. Because finite-size effects are further suppressed once the system size satisfies $L > t^{\frac{1}{z}}$, our choice of $L=10^4$ is sufficiently large. In particular, taking ordinary DP with $z \approx 1.58$, one has $t^{\frac{1}{z}} = 100000^{\frac{1}{1.58}} \approx 1460.69$, and $10^4 \gg 1460.69$.

When $\beta = 1.2$ and $p = 0.2100$, the evolution of particle density is shown by the pink scatter points in Fig.~\ref{f_3}(a), where $\rho(t)$ does not exhibit a global power-law behavior. When $p=0.3100$, the particle density decays more gradually with time and eventually stabilizes at a constant value, deviating from the global power-law characteristic in the critical state. To refine the determination of the critical point, we applied the bisection method to narrow the range of the conditional probability $p$. The results for particle density $\rho(t)$ at additional conditional probabilities are shown in Fig.~\ref{f_3}(a). At the later stages of the finite-size system's time evolution, the differences in initial transfer probabilities become more pronounced, helping to accurately identify the critical region. After discarding situations where the system reaches the absorbing state after a sufficient number of time steps, we can further narrow the critical point range to $p_c\in (0.2568,0.2600)$.

To reduce errors further, we increased the system's average by using $200$ samples of $10^4 \times 10^5$ cluster graphs. At this narrower range of measurements, the results for particle density are shown in Fig.~\ref{f_3}(b). We used a more precise method to quantify the results' reliability and accuracy. As shown in Fig.~\ref{f_3}(b), we obtained the fitting goodness using power-law fitting, defined as:
\begin{equation}
Y^2=1-\frac{\sum\left(y_a-y_p)\right)^2}{\sum\left(y_a-y_m\right)^2},
\end{equation} 
where $y_a$ is the actual value of the particle density, $y_p$ is the corresponding predicted value of the fitted curve, and $y_m$ is the average of the actual values.

\begin{figure*}[t]
\begin{tabular}{cc}
    \includegraphics[width=0.40\textwidth]{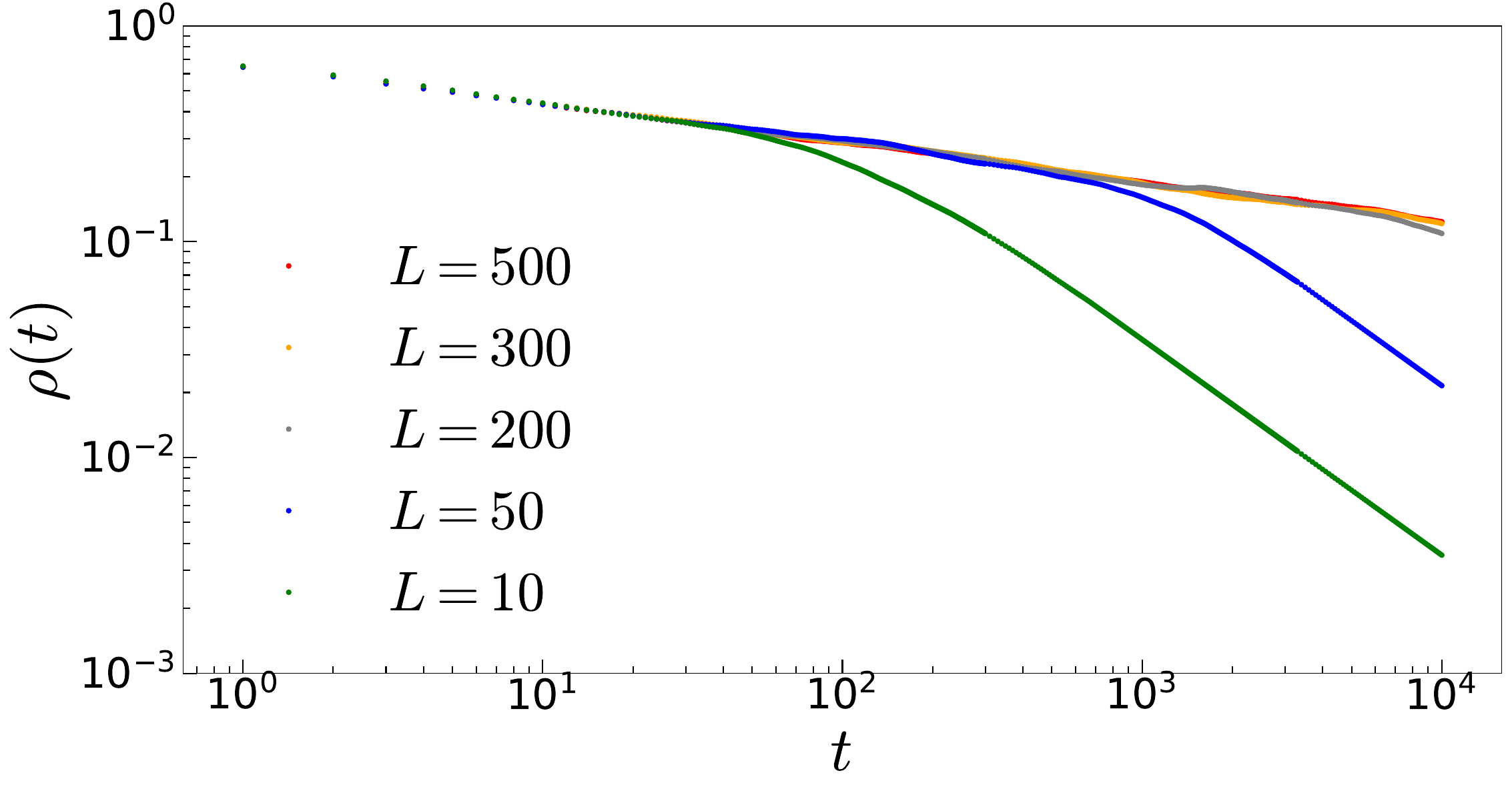} &
    $\qquad$\includegraphics[width=0.40\textwidth]{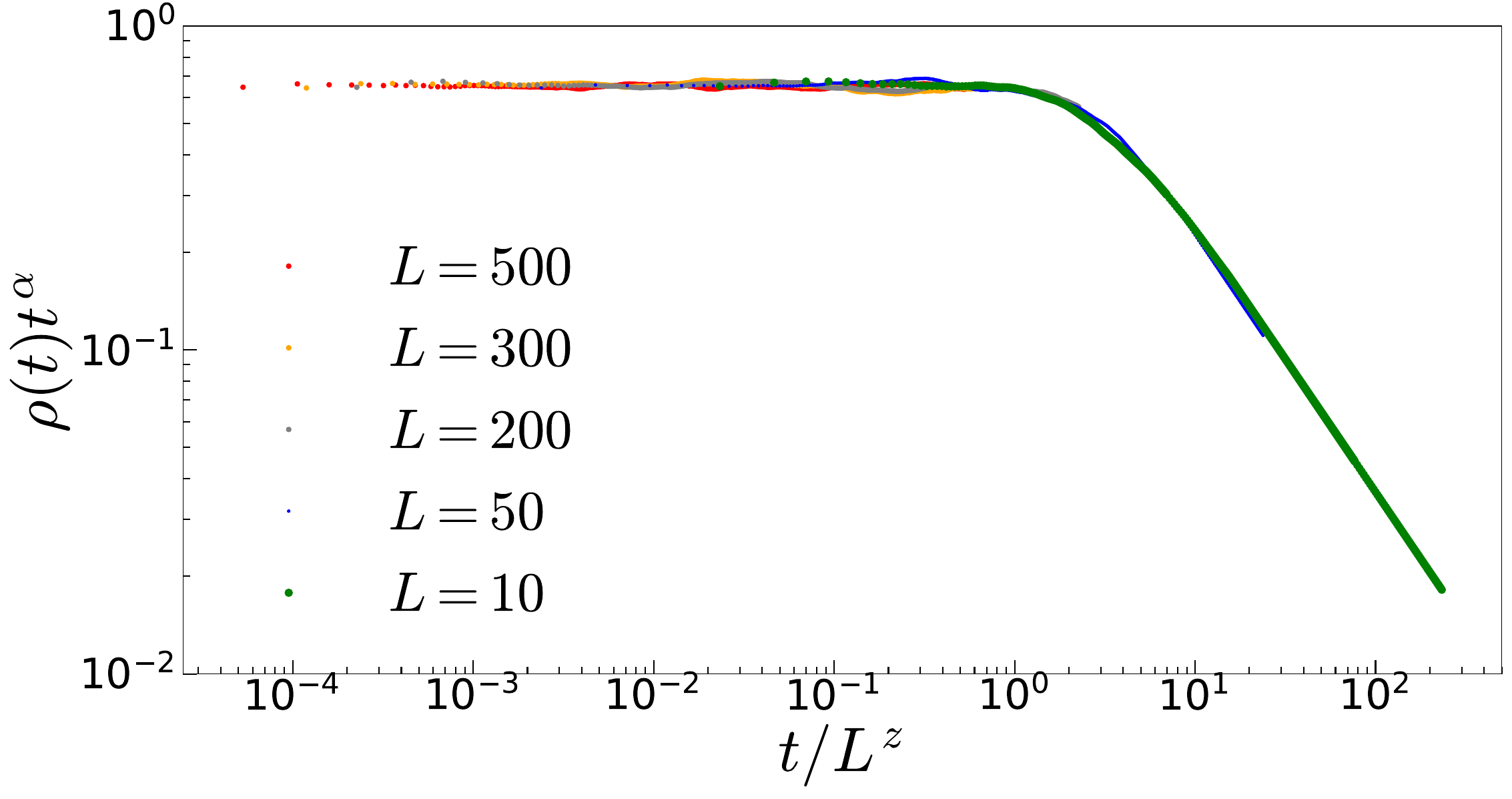} \\
    {\quad}{\quad}(a) & $\qquad$ {\quad}{\quad}(b)
\end{tabular}
\caption{(a)(b) Data collapse of the particle density $\rho(t)$ following the scaling law in (\ref{e_13}) for various system sizes at $\beta=1.2$. The excellent overlap of data across different sizes at $p_c=0.2584$ validates the accuracy of the critical point, allowing for the determination of the dynamic exponent $z=1.58(6)$. With increasing system size, the particle density exhibits more pronounced power-law characteristics. This trend justifies the validity of selecting $L=10^4$ for the simulations.}
\label{f_5}
\end{figure*}

\begin{figure*}[t]
\begin{tabular}{cc}
    \includegraphics[width=0.40\textwidth]{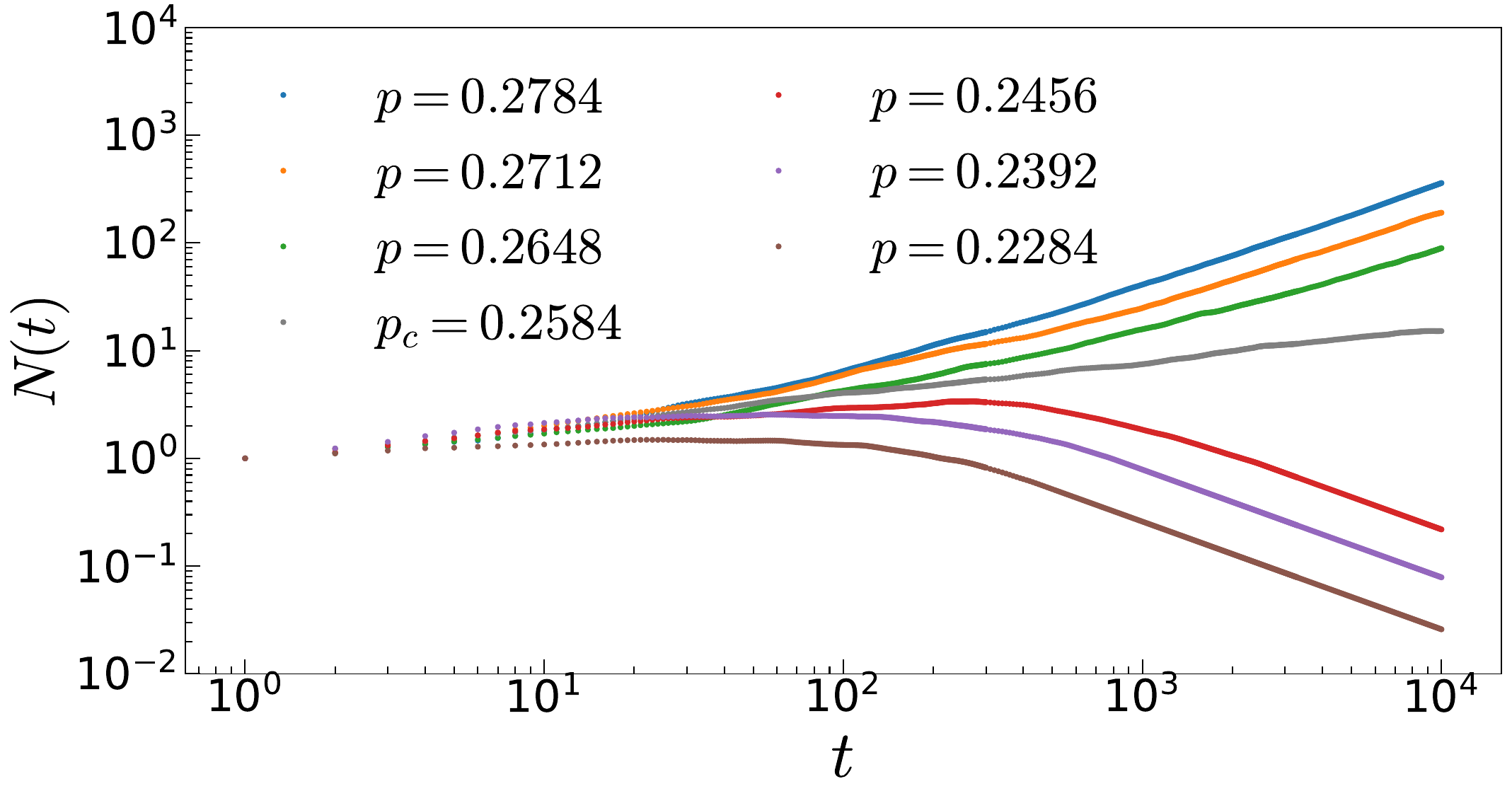} &
    $\qquad$\includegraphics[width=0.40\textwidth]{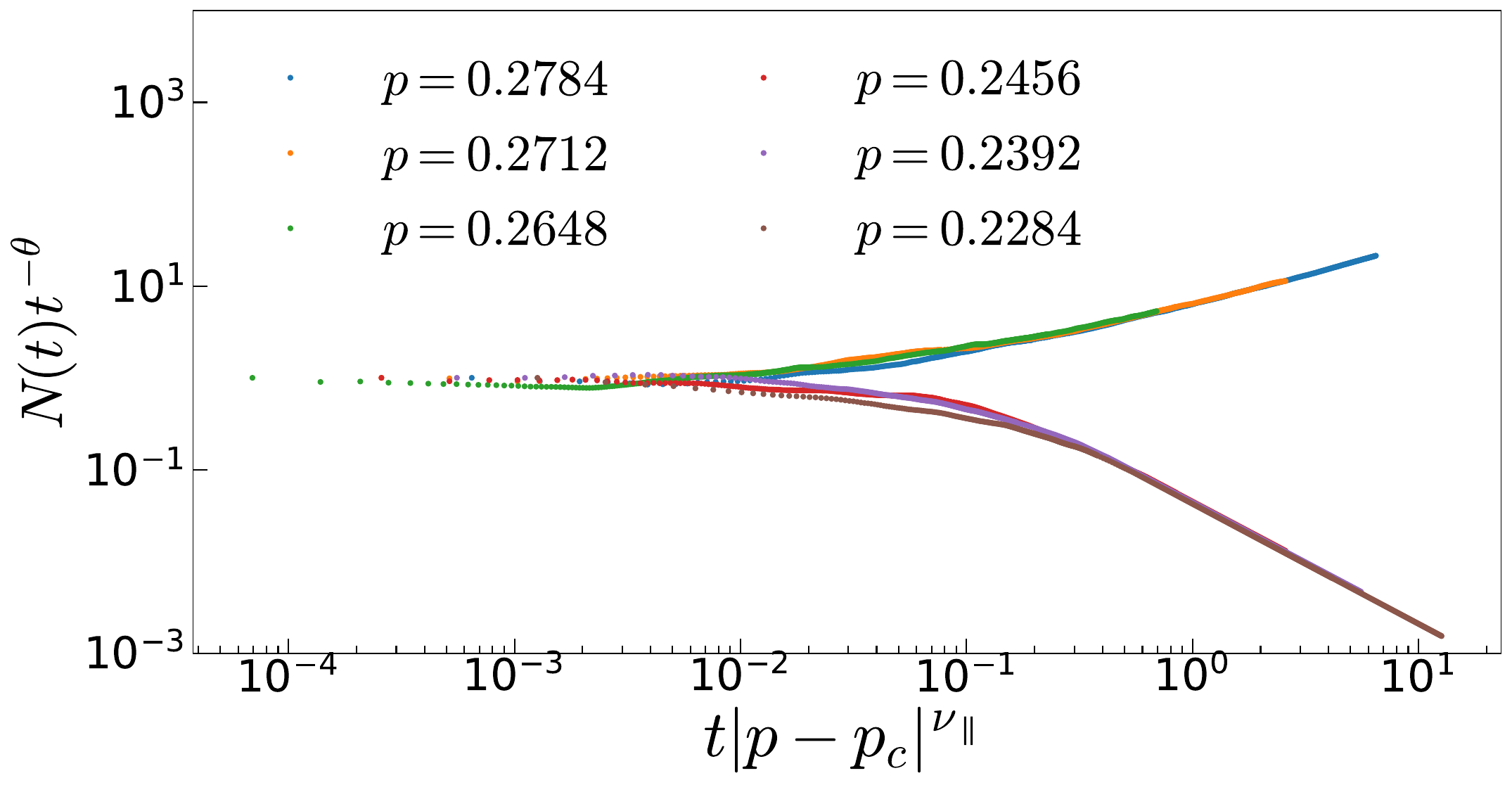} \\
    {\quad}{\quad}(a) & $\qquad$ {\quad}{\quad}(b)
\end{tabular}
\caption{(a)(b) Data collapse plots of the mean particle number $N(t)$ based on the scaling law in \ref{e_15} for varying conditional probabilities at $\beta=1.2$. The effective data collapse near the critical point confirms the accuracy of the critical point and the critical exponent $\theta$. This analysis also directly provides the temporal correlation length critical exponent $\nu_{\parallel}=1.89(6)$}
\label{f_6}
\end{figure*}

The value of $Y^2$ quantifies the goodness of fit for the power-law decay of the particle density $\rho(t)$. For $p=0.2592,0.2588,0.2584,0.2576$, the corresponding $Y^2$ values are $0.9610(8)$,$0.9583(8)$,$0.9946(0)$, and $0.9527(9)$, indicating that the system exhibits optimal power-law decay of particle density at $p=0.2584$. Therefore, when $\beta=1.2$, the critical point of the L\'evy-driven temporally quenched disorder DP system is $p_c=0.258(4)$, with an error smaller than $0.0004$.

The absolute values of fitting residuals are employed as error bars to evaluate the convergence of statistical errors across large data samples, serving as a supplementary analysis of the critical point fitting. The evolution of confidence intervals for a single measurement versus an average of five independent realizations is depicted in Figs.~\ref{f_4}(a) versus (b), based on simulations with a system size of $10^4 \times 10^5$ and 200 independent clusters. With the inclusion of multiple independent measurements, these intervals narrow significantly and exhibit a more uniform distribution around the fitting curve. This behavior confirms the effectiveness of the employed sample size for the precise determination of the critical point.

\subsection{Measurement of the critical exponents $\alpha,\theta$ and $\tilde{z}$}

In this section, we measured the critical exponents $\alpha, \theta, \tilde{z}$ based on the dynamic scaling rate of the absorption phase transition. Furthermore, we validated the accuracy of the critical point and critical exponents using the data collapse method.

In full-seed condition MC simulations, the time evolution of the particle density at the critical point is described by equation(\ref{e_7}). We used this power law to accurately determine the critical point for different values of $\beta$, as $\rho(t)$ exhibits power-law deviations when the conditional probability $p$ deviates from $p_c$. To minimize random errors and accurately determine the critical exponent $\alpha$, we conducted extensive simulations at $\beta = 1.2$ and $p = 0.2584$. As shown in Fig.~\ref{f_4}(b), we generated five sets of $200$ clusters with system sizes of $ 10^4 \times 10^5$ and repeated the simulations $5$ times to average the results. The final measured value of $ \alpha$  was stable at $ \alpha = 0.178(9)$.

The reliability of these measurements is highlighted in Fig.~\ref{a4}(a) of Appendix A, where five independent measurements of $\alpha$ show minimal standard deviations. Complementing this density analysis, the critical point and exponent $\alpha$ are cross-validated via the finite-size scaling law (\ref{e_13}). Fig.~\ref{f_5}(a) and (b) display the temporal evolution of particle density and the corresponding data collapse across different system sizes. A high-quality collapse is achieved by tuning the dynamic exponent to $z=1.58(6)$, verifying the accuracy of the parameters. Additionally, panel (a) clearly visualizes the reduction of finite-size effects in the regime $L > t^{1/z}$.

\begin{figure}
\centering
\includegraphics[width=0.40\textwidth]{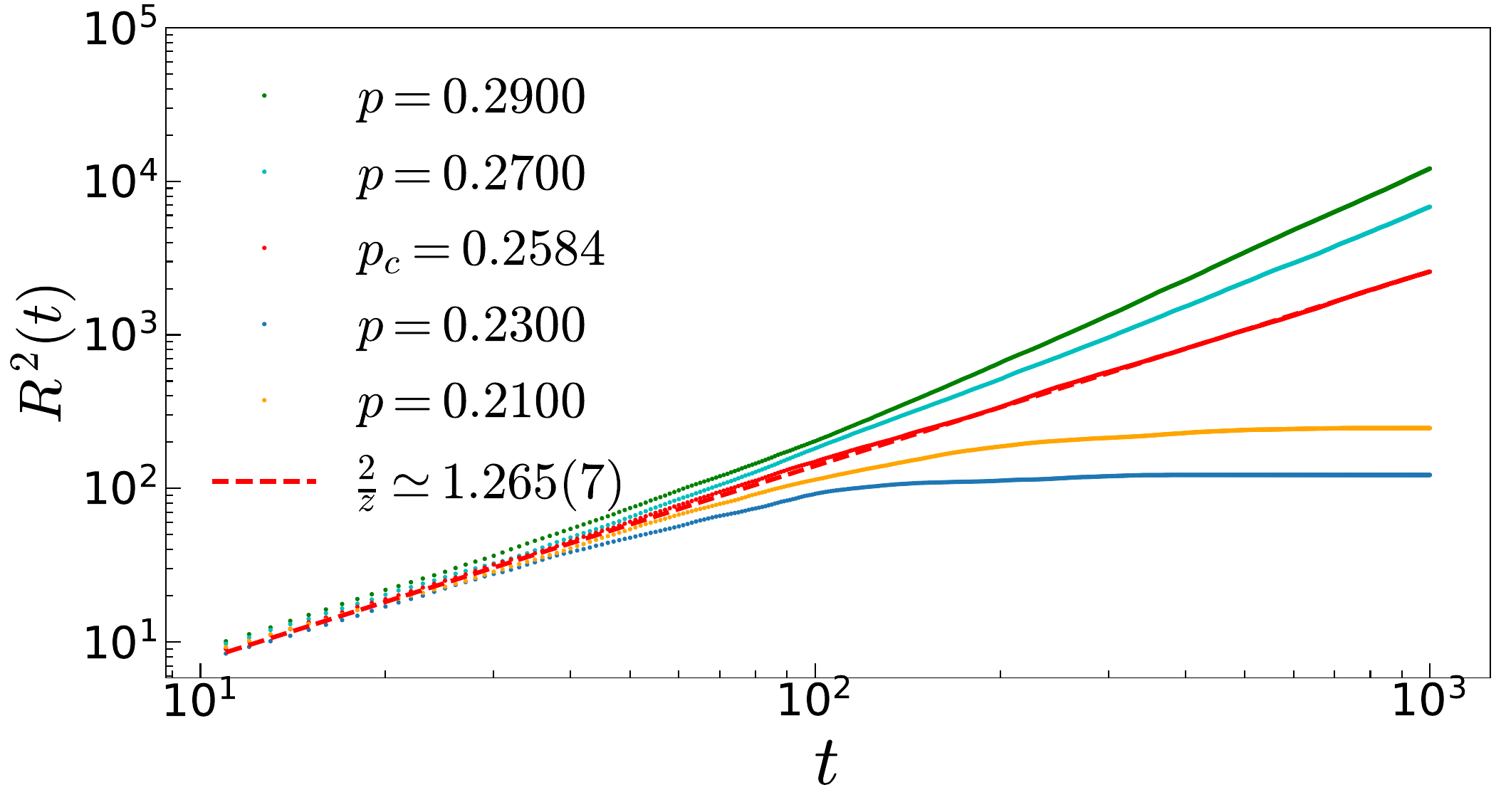}
\caption{For $\beta=1.2$, the value of the critical exponent $\tilde{z}$ is obtained by statistically measuring the mean squared displacement, using the dynamic scaling law (\ref{e_14}).}

\label{f_7}
\end{figure}

\begin{figure*}
\centering
\subfigure[]{\label{fig:subfig:a}
\includegraphics[width=0.38\linewidth]{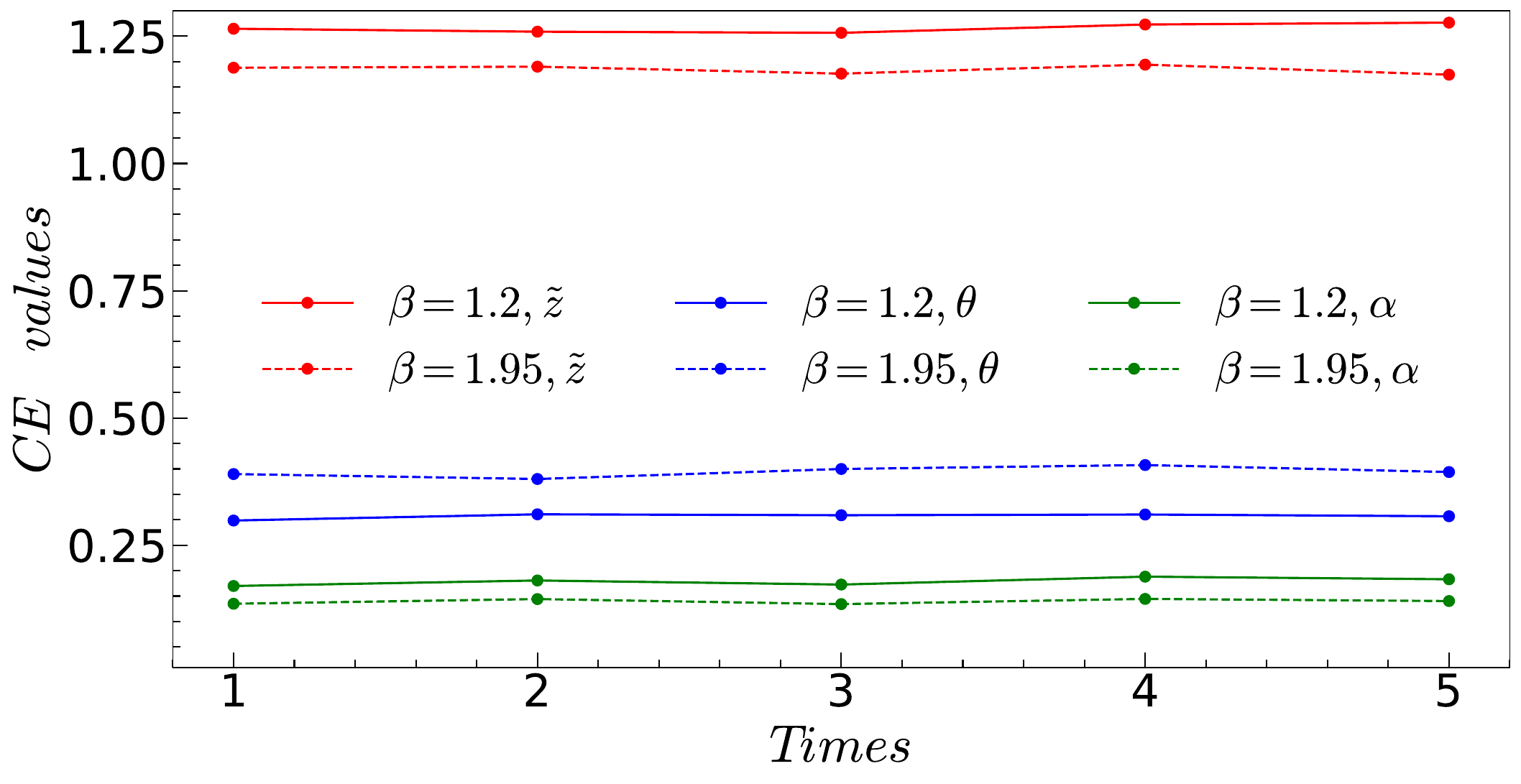}}
\hspace{0.01\linewidth}
\subfigure[]{\label{fig:subfig:b}
\includegraphics[width=0.38\linewidth]{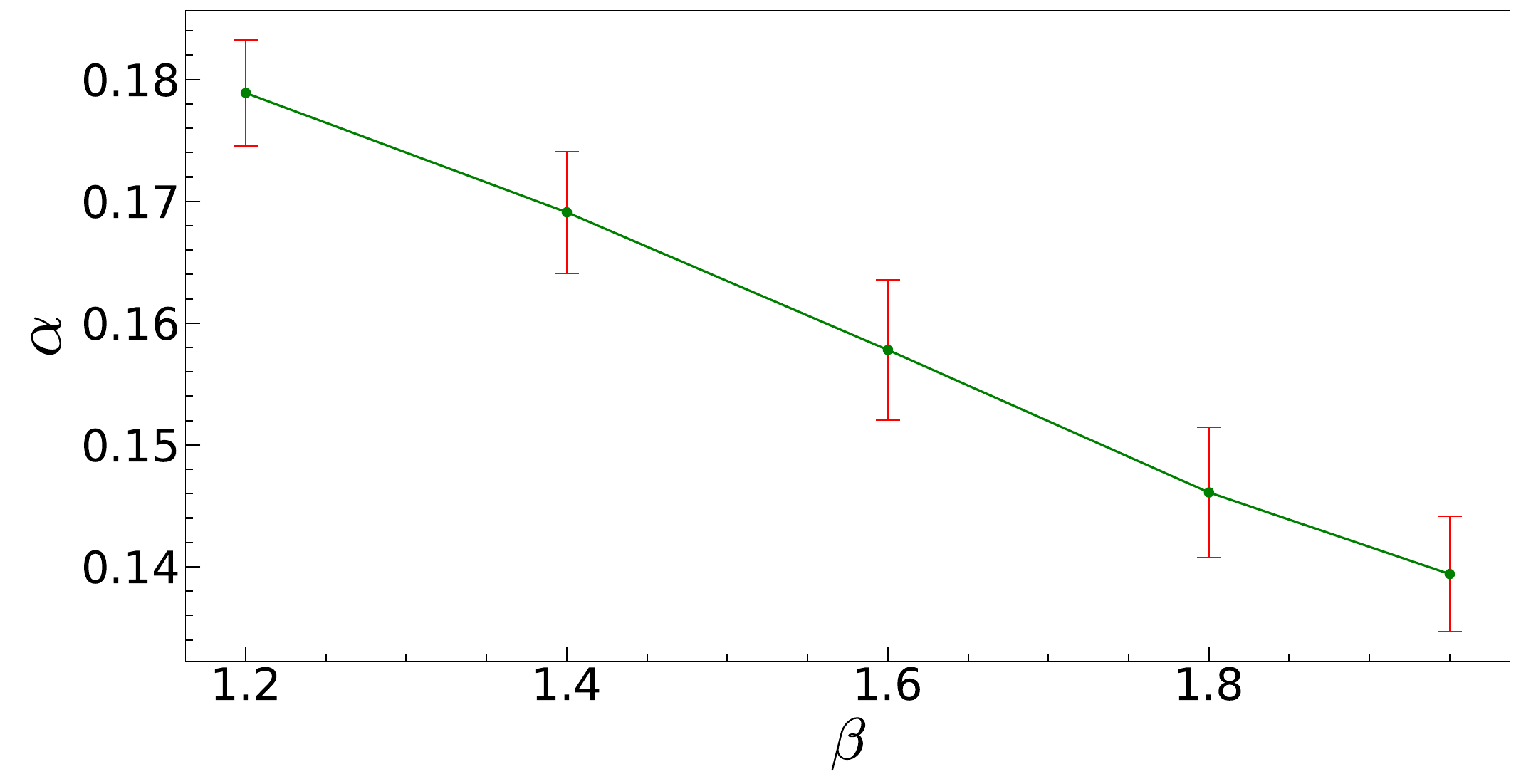}}
\vfill
\subfigure[]{\label{fig:subfig:a}
\includegraphics[width=0.38\linewidth]{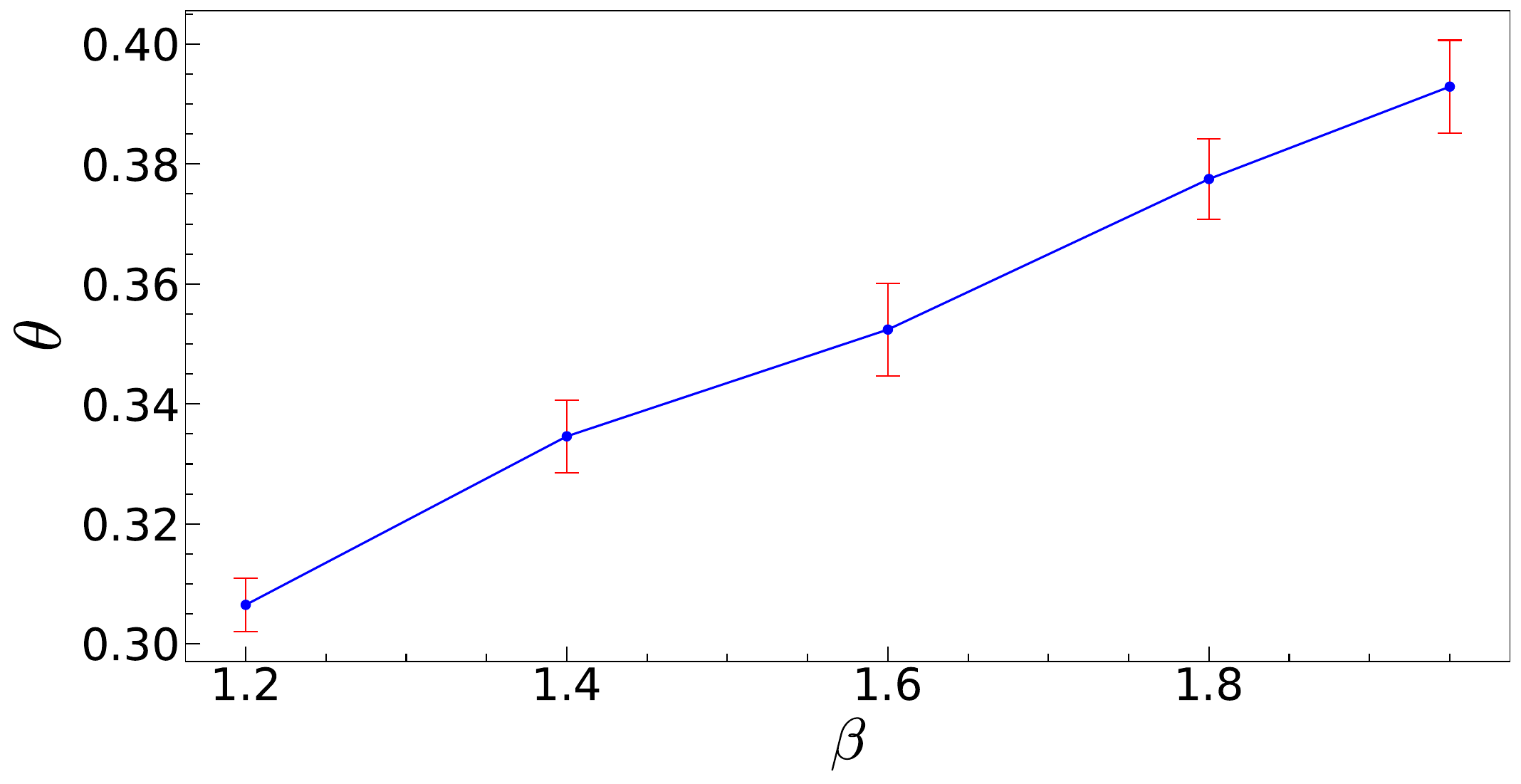}}
\hspace{0.01\linewidth}
\subfigure[]{\label{fig:subfig:b}
\includegraphics[width=0.38\linewidth]{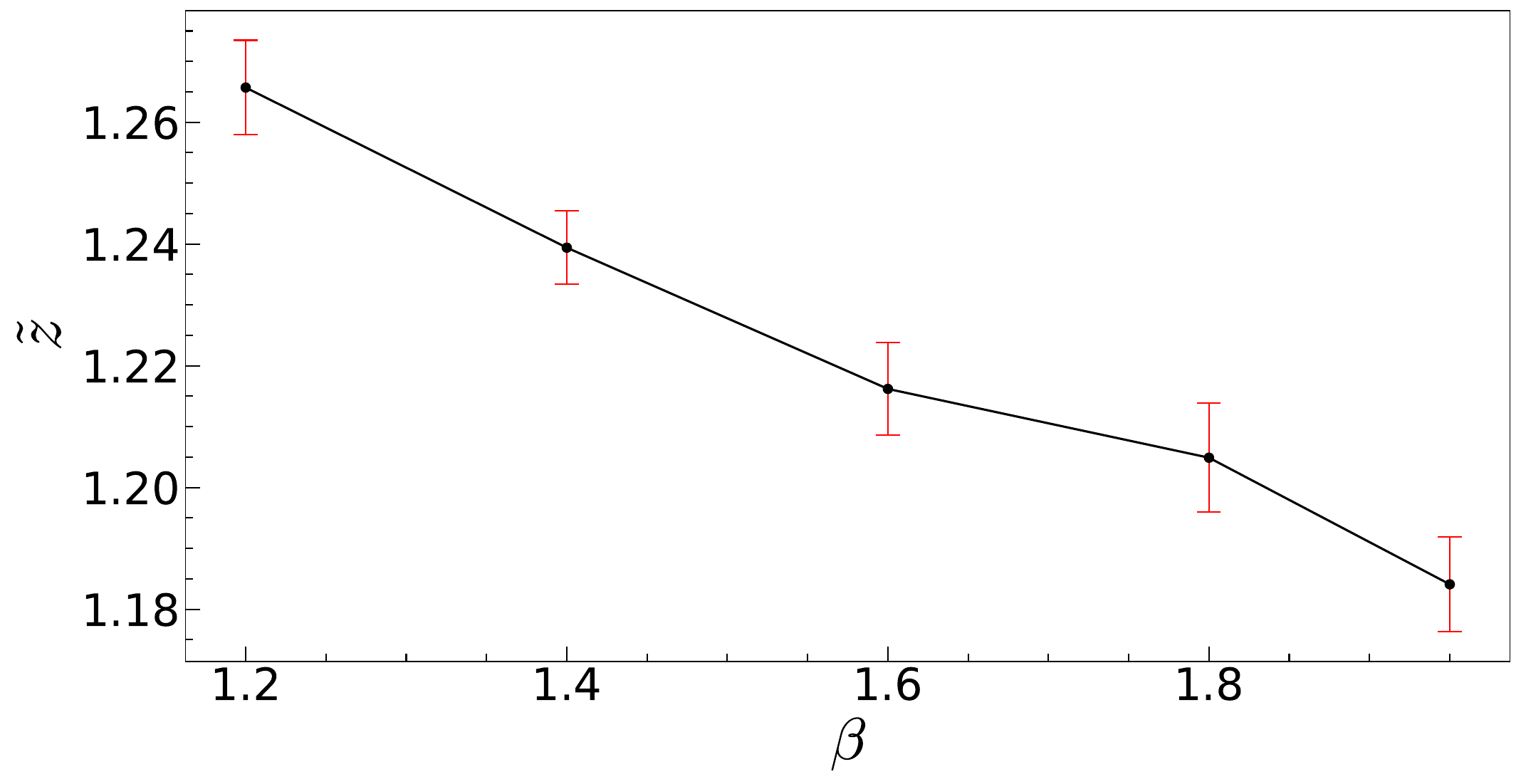}}
\caption{(a)Summary of independent measurements derived from large sample sets. The distribution of five independent realizations for the critical exponents $\alpha$, $\theta$, and $\tilde{z}$ is shown for $\beta=1.2$ and $1.95$. The minimal discrepancies between these independent values confirm the high accuracy of the critical exponents. (b)-(d) Variation of the critical exponents $\alpha$, $\theta$, and $\tilde{z}$ as a function of $\beta$. Error bars represent the standard deviation. These plots clearly illustrate the dependence of the critical exponents on the parameter $\beta$.}
\label{f_8}
\end{figure*}

Unlike the full seed simulation, under one active particle initial conditions, as described in Sec.\uppercase\expandafter{\romannumeral3}, the average total particle number $N(t)$ and mean squared displacement $R^2(t)$ follow the relations(\ref{e_14}). We fixed the initial active seed at the center of the system and used periodic boundary conditions. The sufficiency of the current system size ($L=10^4$) is evidenced by the single-seed cluster growth process shown in Fig.~\ref{a4}(b) of Appendix A, where diffusion fronts rarely encounter the boundaries. Notably, this localized growth structure is a ubiquitous characteristic of the critical state. The preliminary measurement results for the average particle number are shown in Fig.~\ref{f_6}(a), with a system size of $ 10^4 $, $100,000$ time steps, and $2000$ samples for averaging. When $ \beta = 1.2 $ and $ p = 0.2584 $, the fitted value for $ \theta $ was $0.306(5)$. When $p$ deviates from $ 0.2584 $, we observe significant power-law deviations.

The scaling relation (\ref{e_15}) for the mean number of active particles $N(t)$ is used to further verify the critical point and exponent $\theta$. The time evolution of $N(t)$ near the critical point $p_c=0.2584$ for $\beta=1.2$ is shown in Fig.~\ref{f_6}(a).  Subsequently, a data collapse analysis is performed by plotting $N(t)t^{-\theta}$ against the scaling variable $t|p-p_c|^{\nu_{\parallel}}$, as shown in Fig.~\ref{f_6}(b). A good collapse of data points across different $p$ values is achieved by setting the critical exponent to $\nu_{\parallel}=1.89(6)$. This result confirms the high accuracy of the critical point and exponent $\theta$ determined from the large-scale simulations.

Under the same statistical averages, we measured the mean square displacement $R^2(t)$ of the initially active seed, and the power law fitting results are shown in Fig.~\ref{f_7}. Based on the measured values of the exponent $\tilde{z}$ for $\beta=1.2$, we obtain indirect measurements of the dynamic exponent $z$ via the critical exponent relation $\tilde{z}=\frac{2}{z}$. This agrees with the directly measured value of the dynamic exponent $z$ obtained through data collapse under the full-seeds condition. It further illustrates that, in the single-seed simulation environment, the large number of samples we used sufficiently mitigates the influence of finite-size effects. 

The verification analysis of our critical exponent measurements demonstrates the high precision of our sample set within the L\'evy-driven time-quenched disordered DP model.

\subsection{Summary and Analysis of Results}

\begin{table}[!tbp]
	\centering
\resizebox{180pt}{18mm}{	
	\begin{tabular}{ccccc}
			\hline
			\;${\beta}$\; & $p_c$ & $\alpha$ & $\theta$ & $\tilde{z}$\\
			\hline
			\;$1.2$\; & 0.258(4) &0.178(9) & 0.306(5) & 1.265(7) \\
			
                \;$1.4$\; & 0.321(6) & 0.169(1) & 0.334(6) & 1.239(4) \\
			
                \;$1.6$\; & 0.391(4) & 0.157(8) & 0.352(4) & 1.216(2)\\
			
                \;$1.8$\; & 0.487(8) & 0.146(1) & 0.377(5) & 1.204(9)\\
			
                \;$1.95$\; & 0.583(5) & 0.139(4) & 0.392(9)& 1.184(1)\\
			
			\hline
		\end{tabular}
		}

\caption
{The determination of the critical point $p_c$ and the measurements of the critical exponents $\alpha$,  $\theta$, and  $\tilde{z}$ for different values of the parameter  $\beta$.}
\label{table1}
\end{table}

To investigate the influence of the L\'evy distribution parameter $\beta$ on the temporally quenched disordered DP model, we measured the position of the critical point and the values of the critical exponents $\alpha$, $\theta$, and $\tilde{z}$ for various $\beta$ values. Illustrative results for $\beta = 1.95$ are presented in Appendix A. All our results are summarized in Fig.~\ref{f_8} and Table~\ref{table1}.

The measurements of the critical point, as detailed in Table \ref{table1}, reveal that under L\'evy-driven temporally quenched disorder, the DP system indeed exhibits a phase transition with a critical point distinguishing the absorbing state. Larger-scale Monte Carlo simulations lead to a more accurate determination of this critical point. Furthermore, the critical point systematically shifts as the L\'evy distribution parameter $\beta$ is varied.

Five independent measurements of the critical exponents $\alpha$, $\theta$, and $\tilde{z}$ for $\beta=1.2$ and $1.95$ are shown in Fig.~\ref{f_8}(a). The minimal differences observed between these measurements indicate the high statistical reliability of the average results. Furthermore, the variation of these critical exponents with respect to $\beta$ is plotted in Figs. \ref{f_8}(b)-(d), where error bars represent the standard deviation. These results clearly show that the critical exponents depend strongly on the parameter $\beta$.

From the critical exponent $\alpha$ measurements in Fig.~\ref{f_8} (b), we observe that as the L\'evy distribution parameter $\beta$ increases, $\alpha$ decreases. This is consistent with the characteristics of the L\'evy distribution: as $\beta$ increases, the kurtosis decreases, and the distribution becomes flatter. This suggests that the quenching of the conditional probability becomes less severe at higher $\beta$, causing the particle reaction-diffusion process to proceed more slowly over time, leading to a gradual change in particle density. The observed trend in the variation of the critical exponents with the distributional characteristics supports the physical interpretation.

For the average number of surviving particles, under single-seed conditions, as $\beta$ increases, the increment in the probability of a flatter distribution leads to more particles undergoing branching. Therefore, in terms of long-term evolution, the critical exponent $\theta$ becomes larger. Considering the increase in the dynamic exponent $z$ from the perspective of spatial correlation length, due to the occurrence of edge avalanche effects, large vacancies appear in the cluster growth of the temporally quenched disorder DP system, and the cluster boundaries further expand into space. The cluster structure tends to be dispersed, thus the enhancement of spatial correlation leads to an increase in the dynamic exponent $z$, i.e., a decrease in the spreading exponent $\tilde{z}$. 

The measurements of the critical exponents $\alpha$, $\theta$, and $\tilde{z}$ indicate that under the control of $\beta$, the evolutionary characteristics of the L\'evy-driven temporally quenched disorder DP system continuously change.

\section{Conclusion}

In this paper, we investigated the absorbing phase transition behavior of a temporally quenched disordered DP system driven by the CDF of the L\'evy distribution. Through the analysis of the system's phase diagram and properties near the critical point, we confirmed the scaling characteristics of this model. Adhering to the dynamic scaling theory of absorbing phase transitions, we determined the location of the system's critical point, regulated by the control parameter $\beta$ associated with different L\'evy distributions. Our results indicate that while temporally quenched disorder modifies the critical region of an ordinary DP system, it preserves the power-law decay behavior of particle density.

To enhance the accuracy of the critical point and critical exponents, we employed a multi-faceted approach. This included robust statistical judgments of goodness-of-fit, averaging measurements from a large number of independent sample sets, and data collapse under the framework of scaling theory. Consequently, by observing the critical exponents $\alpha, \theta,$ and $\tilde{z}$, we found that significant variations in $\beta$ lead to notable changes in the system's critical exponents. The intensity of the quenched disorder influences the spatiotemporal symmetry of the DP system and can potentially cause the system to deviate from the DP universality class. Unlike uniform quenched disorder, adjusting parameters such as the kurtosis and variance of the L\'evy distribution may offer a more accurate mapping to real physical systems.

When adjusting the parameters of a stable L\'evy distribution, it is possible to map the general L\'evy distribution to more common distributions such as the Gaussian or Cauchy distributions. From the perspective of methodological extensibility, temporally quenched disorder driven by L\'evy distributions offers broad application potential in the study of absorbing phase transitions. Given that the implementation of random step generation programs satisfying L\'evy distributions can incur increased computational costs, simpler distributions like the uniform or Gaussian distribution can be considered as alternative options for reaction-diffusion systems with less complex quenching mechanisms. The analytical framework, which preserves the general characteristics of absorbing phase transitions, suggests that spatiotemporal quenched disorder driven by specific distributions holds promising theoretical and experimental prospects across a variety of reaction-diffusion processes.

\section{Acknowledgements}
This work is supported in part by the National Key Research and Development Program of China under Grant No. 2024YFA1611003, the Fundamental Research Funds for Central China Normal University(CCNU24JC007), and the 111 Project, with Grant No. BP0820038. Financially supported by self-determined research funds of CCNU from the college basic research and operation of MOE.

\appendix
\section{Supplementary Numerical Simulation Results}

In this appendix, we supplement the results of some numerical simulations conducted in this work, which were not fully presented in the main text. The main content includes illustrative examples of the results for Levy-distributed driven time-quenched disorder in DP at different $\beta$ values. In addition, there is the error analysis of the critical exponent $\alpha$ mentioned in the main text, and the cluster configuration analysis of time-quenched disordered DP on the system sizes at which we obtain the observable values.

We supplement the measurements of the critical point and critical exponents for $\beta = 1.95$. Fig.~\ref{a1} displays the time evolution of particle density under different conditional probabilities, along with the stability of the confidence intervals obtained from repeated measurements under optimal fitting conditions. Fig.~\ref{a2} shows the data collapse results obtained using finite-size scaling analysis, and Fig.~\ref{a3} presents the data collapse of the average particle number under different conditional probabilities. These results provide strong evidence for the accuracy of our critical exponent measurements.

\setcounter{figure}{0}
\renewcommand{\thefigure}{A\arabic{figure}}

\begin{figure*}
\begin{tabular}{cc}
    \includegraphics[width=0.38\textwidth]{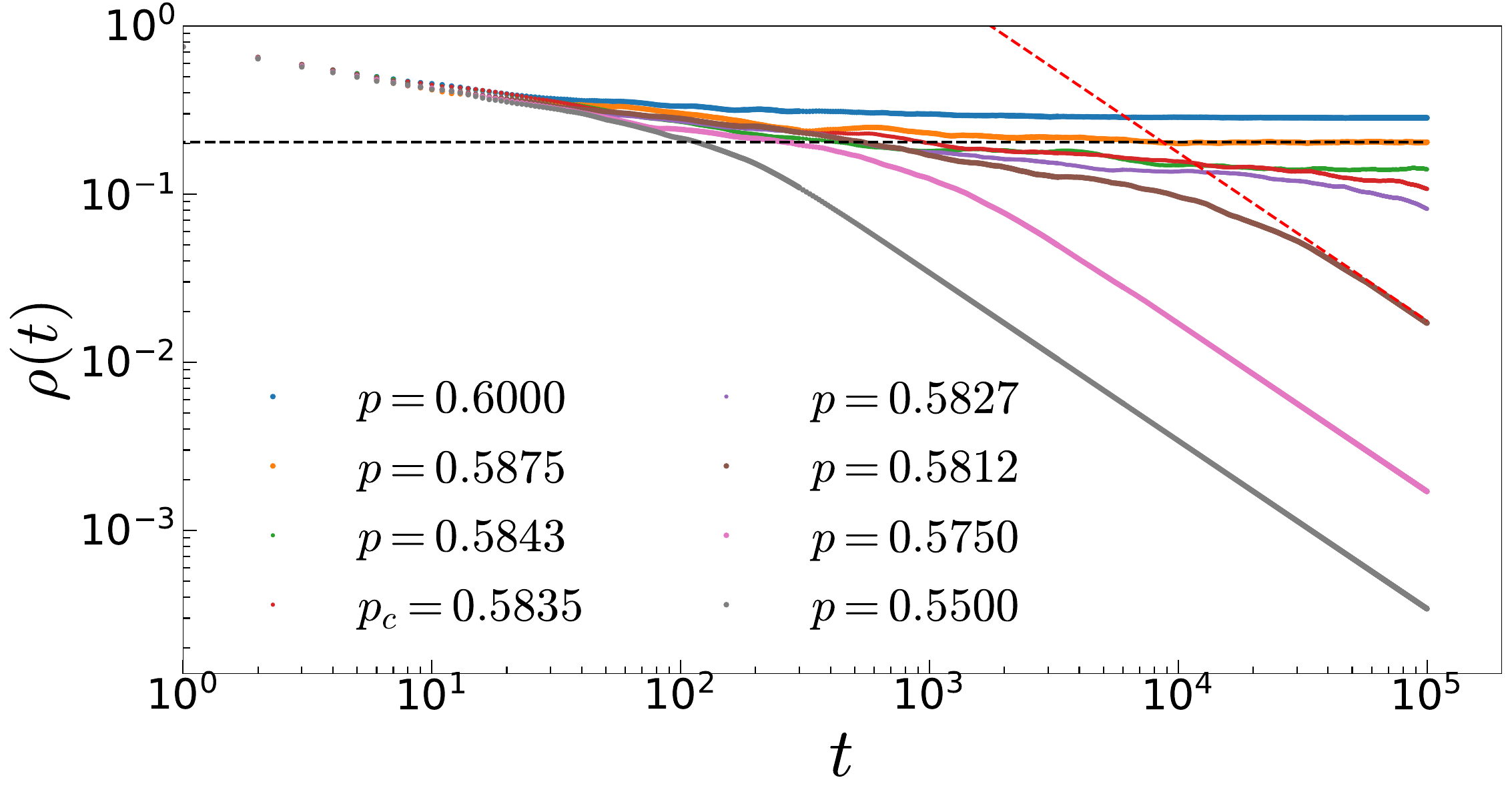} &
    $\qquad$\includegraphics[width=0.38\textwidth]{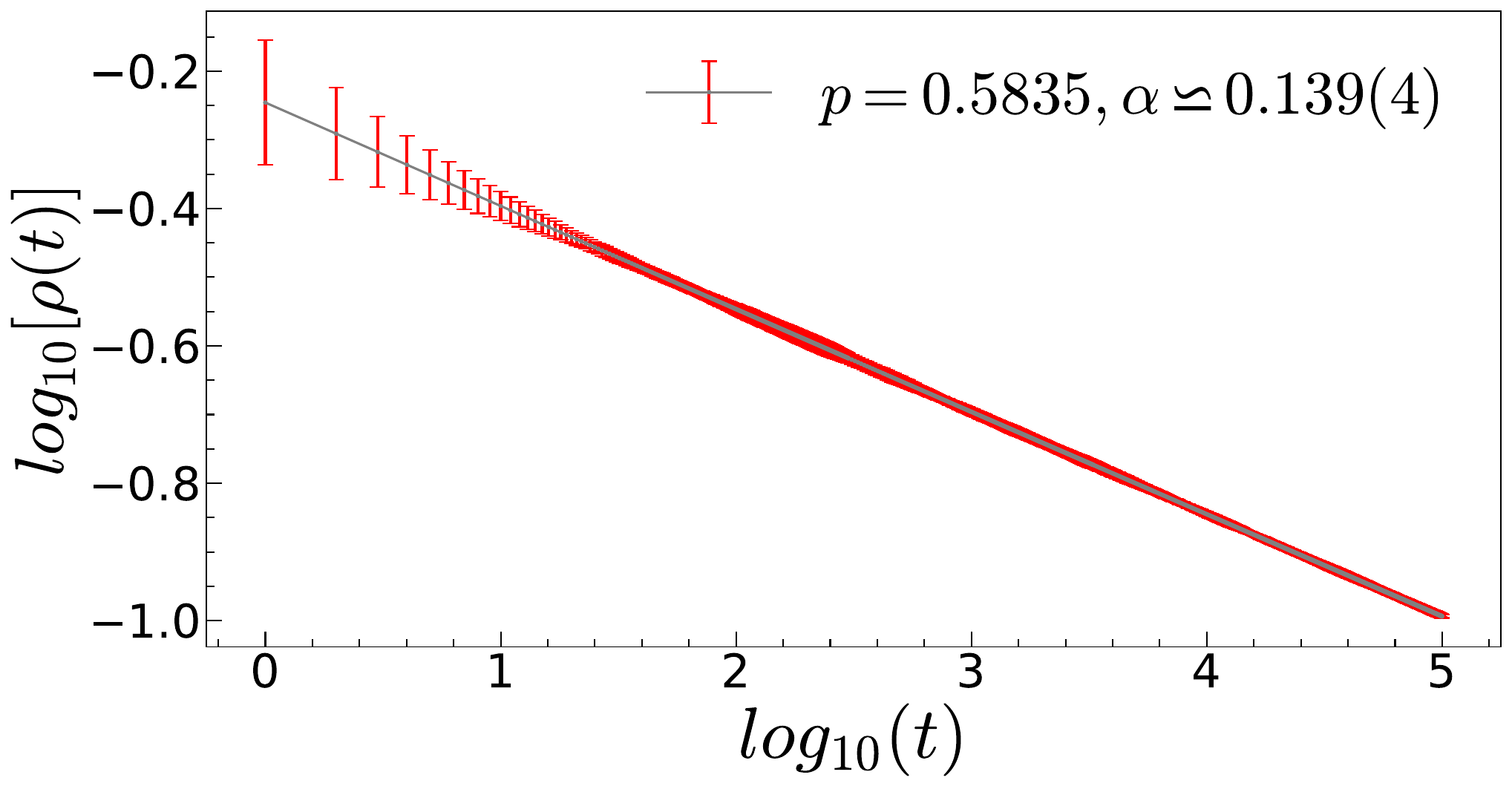} \\
    {\quad}{\quad}(a) & $\qquad$ {\quad}{\quad}(b)
\end{tabular}
\caption{Determination of the critical point at $\beta = 1.95$. (a) Time evolution of the particle density under various conditional probabilities. (b) Stability of the confidence intervals derived from repeated measurements under best-fit conditions.}
\label{a1}
\end{figure*}

\begin{figure*}[htbp]
\begin{tabular}{cc}
    \includegraphics[width=0.38\textwidth]{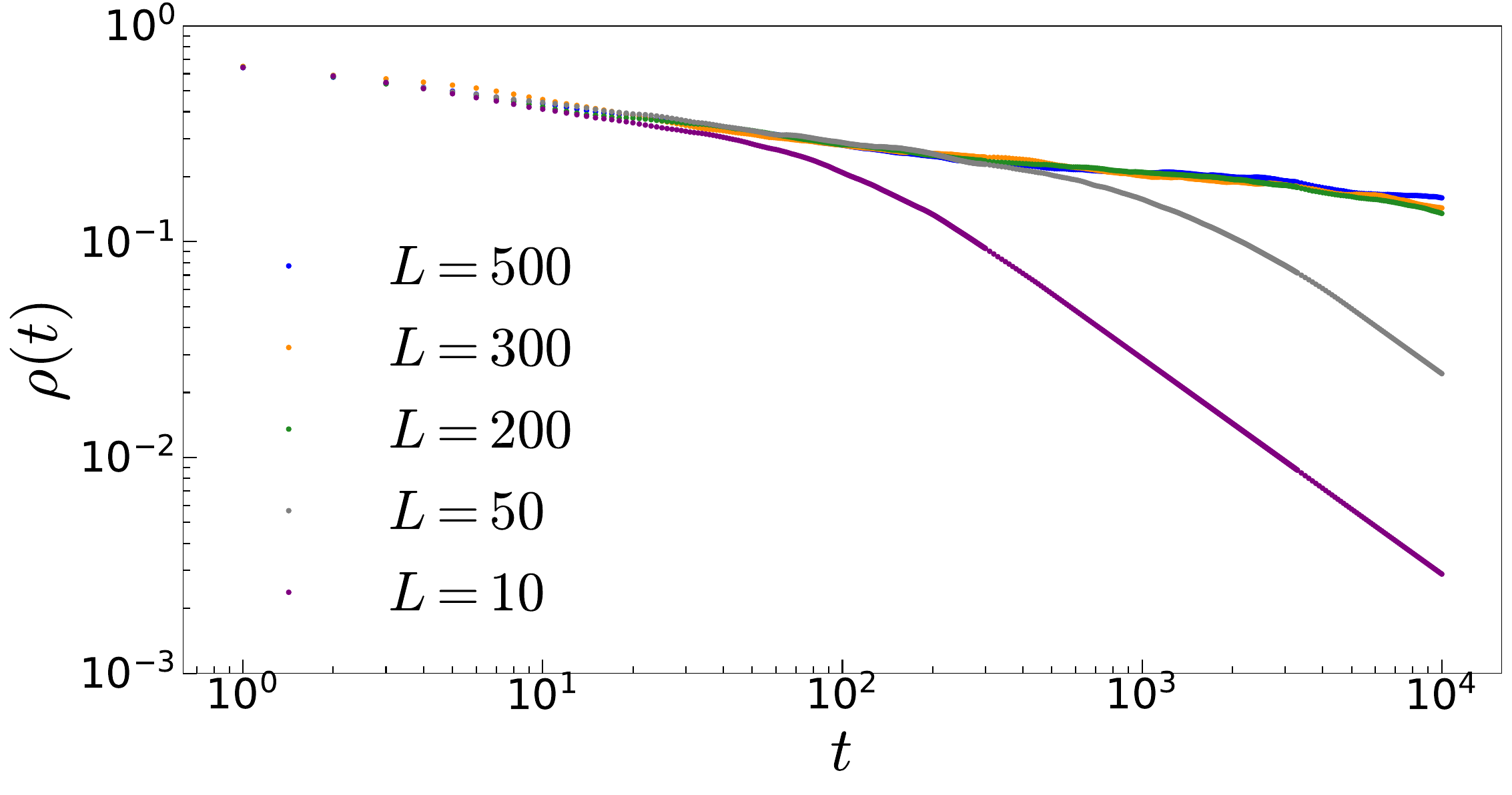} &
    $\qquad$\includegraphics[width=0.38\textwidth]{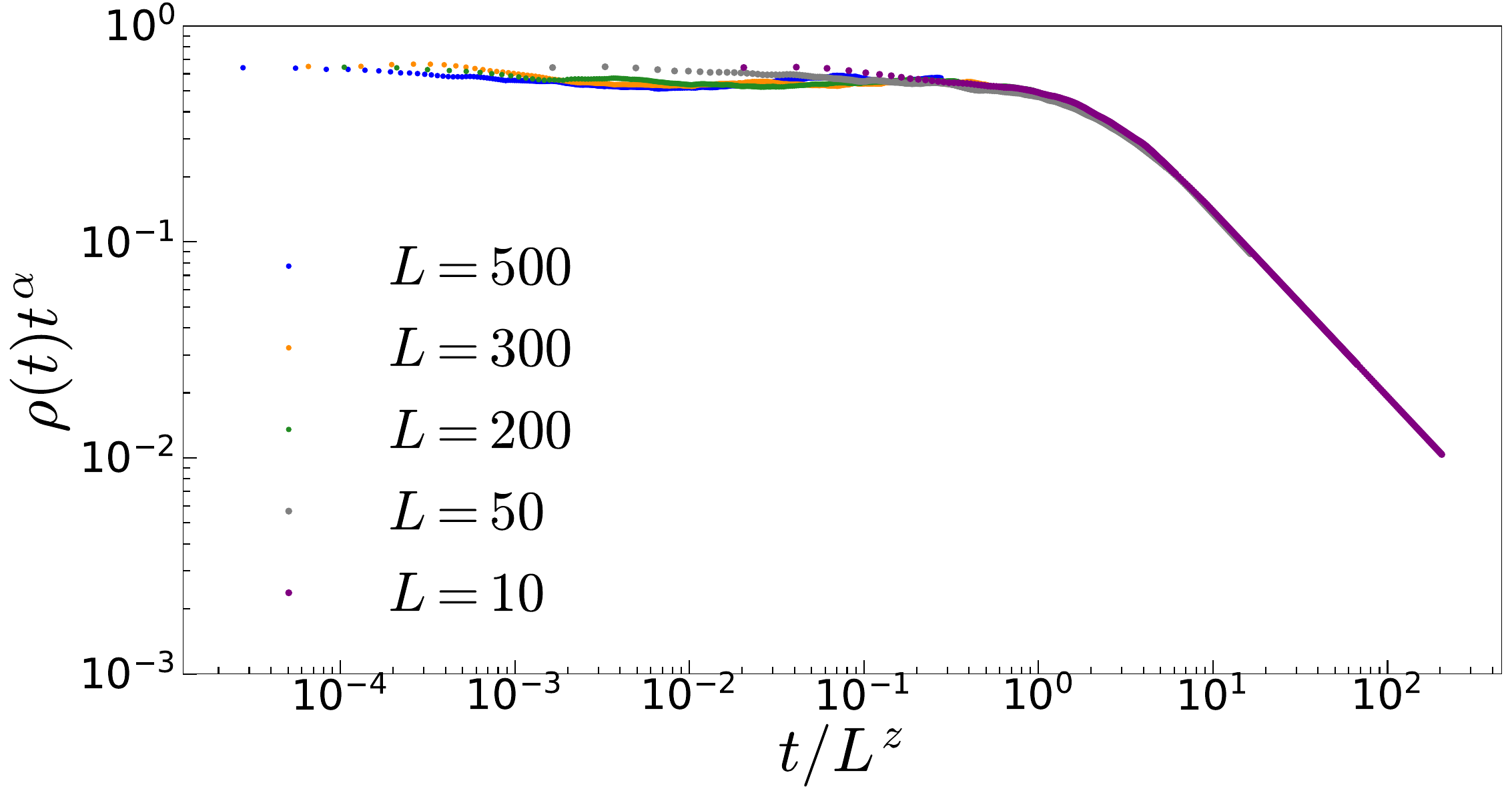} \\
    {\quad}{\quad}(a) & $\qquad$ {\quad}{\quad}(b)
\end{tabular}
\caption{Examples of the analysis at $\beta=1.95$ ($p_c=0.5835$). A high-quality data collapse is achieved by tuning the dynamic exponent to $z=1.69(5)$. This result further corroborates the reliability of the determined critical point and exponents under this parameter setting.}
\label{a2}
\end{figure*}

\begin{figure*}[htbp]
\begin{tabular}{cc}
    \includegraphics[width=0.38\textwidth]{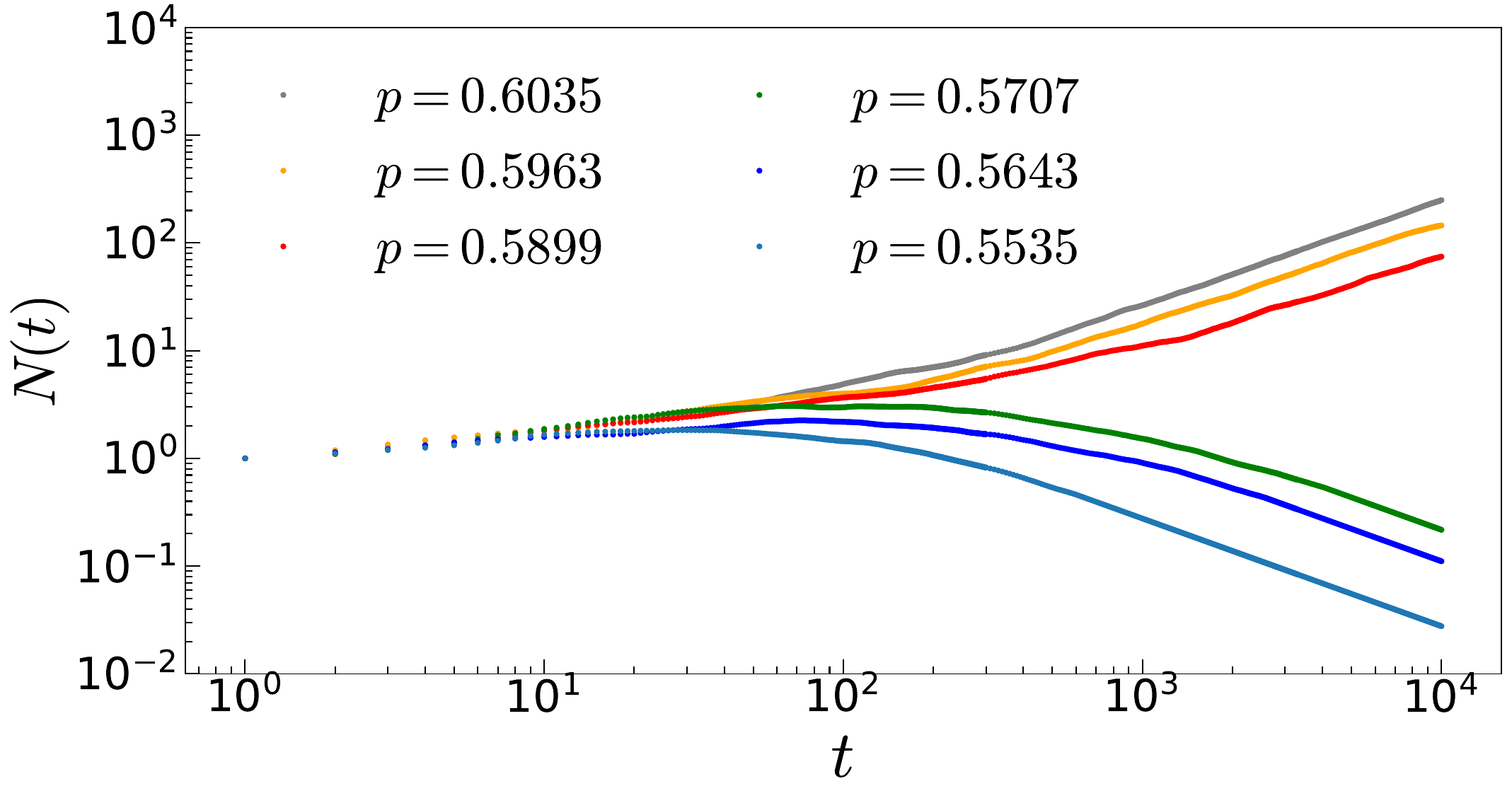} &
    $\qquad$\includegraphics[width=0.38\textwidth]{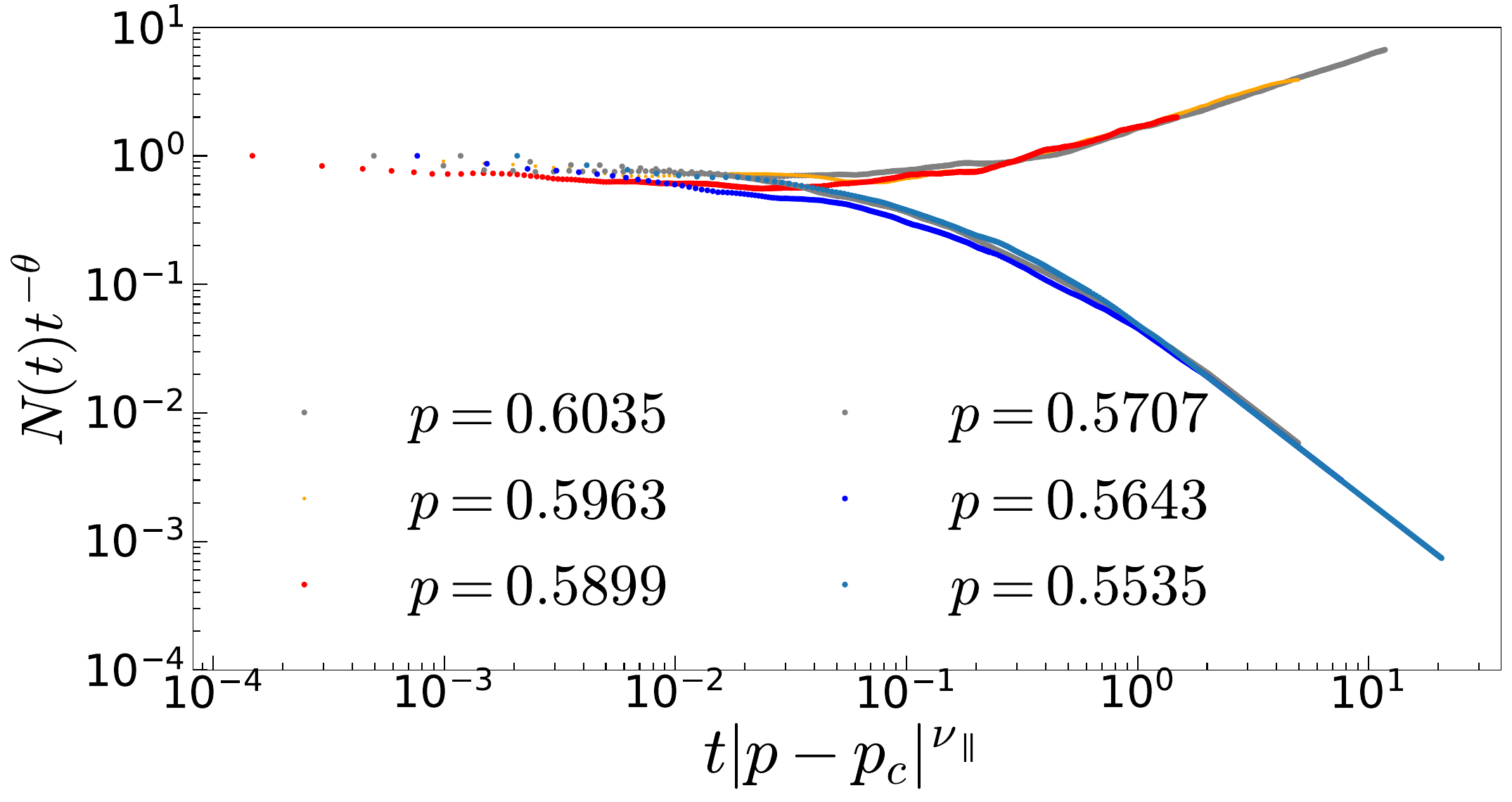} \\
    {\quad}{\quad}(a) & $\qquad$ {\quad}{\quad}(b)
\end{tabular}
\caption{Examples of the scaling analysis for $\beta=1.95$. The good data collapse obtained by setting $\nu_{\parallel}=1.74(9)$ indicates the accuracy of the critical point and exponent $\theta$ under the variation of parameter $\beta$.}
\label{a3}
\end{figure*}

\begin{figure*}[htbp]
\begin{tabular}{cc}
    \includegraphics[width=0.38\textwidth]{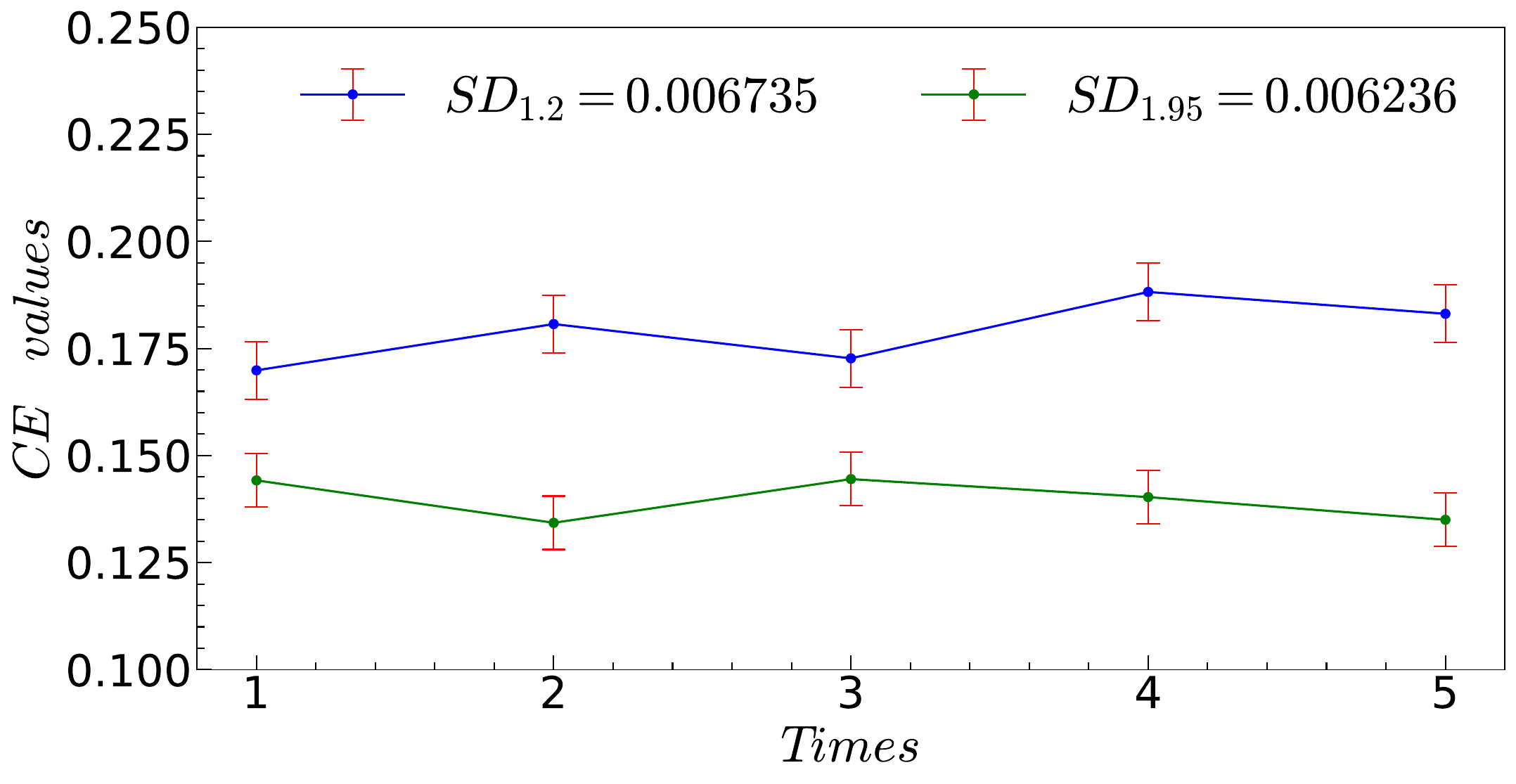} &
    $\qquad${\quad}{\quad}{\quad}{\quad}\includegraphics[width=0.19\textwidth]{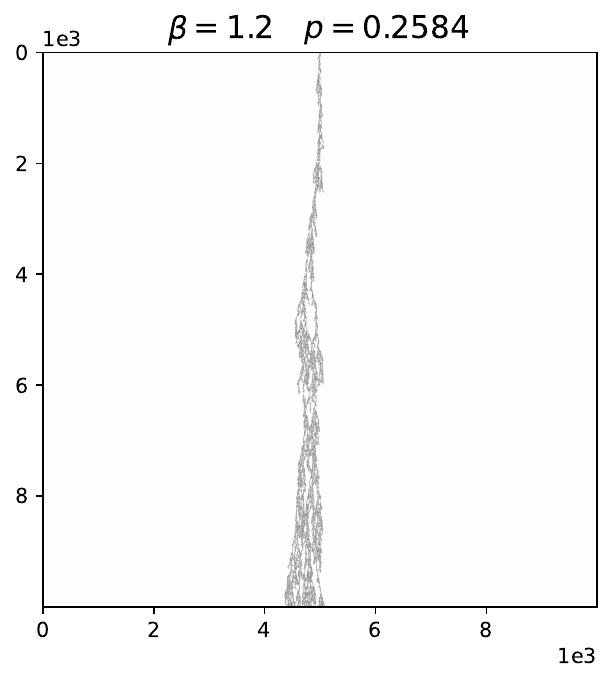} \\
    {\quad}{\quad}(a) & $\qquad$ {\quad}{\quad}(b)
\end{tabular}
\caption{(a) Results of five independent measurements of the critical exponent $\alpha$ at $\beta = 1.2$ and $1.95$, where the legend indicates error bars determined by the standard deviation. The ensemble-averaged results yield corresponding $\alpha$ values of $0.178(9)$ and $0.139(4)$, respectively. (b) An instance of critical point cluster growth structure under single-seed conditions. Such evolutionary structures, commonly observed near the critical point and typically unable to reach the system boundaries in its vicinity, effectively demonstrate that the selected system size can significantly mitigate the impact of finite-size effects.}
\label{a4}
\end{figure*}

The standard deviations of the critically measured critical exponent $\alpha$ at $\beta=1.2$ and $1.95$ are structurally presented in Fig.~\ref{a4}(a), which illustrates the strong dependence of the critical exponent on the distribution parameters. As stated in the main text, system sizes of $10^4 \times 10^4$ are sufficient to mitigate the effects of finite-size scaling. The reduction in finite-size effects is clearly visible in Fig.~\ref{f_5}(a) of the main text, where at the critical point, increasing the system size leads to the dynamic evolution of particle density approaching power-law behavior. Furthermore, the general cluster structure of the time-quenched disordered DP system at the critical point, as shown in Fig.~\ref{a4}(b), further supports this.

\nocite{*}

\bibliography{apssamp}

@PREAMBLE{
 "\providecommand{\noopsort}[1]{}" 
 # "\providecommand{\singleletter}[1]{#1}%" 
}

@book{christensen2005complexity,
  title={{Complexity and criticality}},
  author={Christensen, Kim and Moloney, Nicholas R},
  volume={1},
  year={2005},
  publisher={World Scientific Publishing Company}
}

@book{henkel2008non,
  title={Non-equilibrium phase transitions},
  author={Henkel, Malte and Hinrichsen, Haye and L{\"u}beck, Sven and Pleimling, Michel},
  volume={1},
  year={2008},
  publisher={Springer}
}

@article{lubeck2006crossover,
  title={Crossover scaling in the Domany--Kinzel cellular automaton},
  author={L{\"u}beck, S},
  journal={Journal of Statistical Mechanics: Theory and Experiment},
  volume={2006},
  number={09},
  pages={P09009},
  year={2006},
  publisher={IOP Publishing},
  note={doi:{\color{blue}\href{http://dx.doi.org/10.1088/1742-5468/2006/09/p09009}{10.1088/1742-5468/2006/09/p09009}}}
}

@article{ma2024emergent,
  title={Emergent topological ordered phase for the Ising-XY model revealed by cluster-updating Monte Carlo method},
  author={Ma, Heyang and Zhang, Wanzhou and Tian, Yanting and Ding, Chengxiang and Deng, Youjin},
  journal={Chinese Physics B},
  volume={33},
  number={4},
  pages={040503},
  year={2024},
  publisher={IOP Publishing},
 note={doi:{\color{blue}\href{http://dx.doi.org/10.1088/1674-1056/ad1d4d}{10.1088/1674-1056/ad1d4d}}}

}

@article{deng2024chimera,
  title={Chimera-like states in neural networks and power systems},
  author={Deng, Shengfeng and {\'O}dor, G{\'e}za},
  journal={Chaos: An Interdisciplinary Journal of Nonlinear Science},
  volume={34},
  number={3},
  year={2024},
  publisher={AIP Publishing},
note={doi:{\color{blue}\href{http://dx.doi.org/10.1063/5.0154581}{10.1063/5.0154581}}}

}

@article{hinrichsen2000non,
  title={Non-equilibrium critical phenomena and phase transitions into absorbing states},
  author={Hinrichsen, Haye},
  journal={Advances in physics},
  volume={49},
  number={7},
  pages={815--958},
  year={2000},
  publisher={Taylor \& Francis},
  note={doi:{\color{blue}\href{http://dx.doi.org/10.1080/00018730050198152}{10.1080/00018730050198152}}}
}

@article{grassberger1979reggeon,
  title={Reggeon field theory (Schl{\"o}gl's first model) on a lattice: Monte Carlo calculations of critical behaviour},
  author={Grassberger, Peter and de la Torre, Alberto},
  journal={Annals of Physics},
  volume={122},
  number={2},
  pages={373--396},
  year={1979},
  publisher={Elsevier},
  note={doi:{\color{blue}\href{http://dx.doi.org/10.1016/0003-4916(79)90109-X}{10.1016/0003-4916(79)90109-X}}}
}

@article{janssen1981nonequilibrium,
  title={On the nonequilibrium phase transition in reaction-diffusion systems with an absorbing stationary state},
  author={Janssen, Hans-Karl},
  journal={Zeitschrift f{\"u}r Physik B Condensed Matter},
  volume={42},
  number={2},
  pages={151--154},
  year={1981},
  publisher={Springer},
note={doi:{\color{blue}\href{http://dx.doi.org/10.1007/BF01319549}{10.1007/BF01319549}}}
}

@article{1982On,
  title={On phase transitions in Schlgl's second model},
  author={ Grassberger, P. },
  journal={Zeitschrift für Physik B Condensed Matter},
  volume={47},
  number={4},
  pages={365-374},
  year={1982},
  note={doi:{\color{blue}\href{http://dx.doi.org/10.1007/BF01313803}{10.1007/BF01313803}}}
}

@article{1984Equivalence,
  title={Equivalence of Cellular Automata to Ising Models and Directed Percolation},
  author={ Domany, E.  and  Kinzel, W. },
  journal={Physical Review Letters},
  volume={53},
  number={4},
  pages={311-314},
  year={1984},
  note={doi:{\color{blue}\href{http://dx.doi.org/10.1103/PhysRevLett.53.311}{10.1103/PhysRevLett.53.311}}}
}

@article{1985Phase,
  title={Phase transitions of cellular automata},
  author={ Kinzel, W. },
  journal={Zeitschrift Für Physik B Condensed Matter},
  volume={58},
  number={3},
  pages={229-244},
  year={1985},
  note={doi:{\color{blue}\href{http://dx.doi.org/10.1007/BF01309255}{10.1007/BF01309255}}}
}

@book{2005field,
  title={Field theory, the renormalization group, and critical phenomena: graphs to computers},
  author={Amit, Daniel J and Martin-Mayor, Victor},
  year={2005},
  publisher={World Scientific Publishing Company}
}

@article{zhong1995universality,
  title={Universality class of two-offspring branching annihilating random walks},
  author={Zhong, Dexin and ben-Avraham, Daniel},
  journal={Physics Letters A},
  volume={209},
  number={5-6},
  pages={333--337},
  year={1995},
  publisher={Elsevier},
  note={doi:{\color{blue}\href{http://dx.doi.org/10.1016/0375-9601(95)00869-1}{10.1016/0375-9601(95)00869-1}}}
}

@misc{whitney2025generalapproachstatisticsmicrobial,
      title={A general approach to the statistics of microbial orientation: L\'{e}vy walks, noise, and deterministic motion}, 
      author={Taylor Whitney and Thomas Solomon and Kevin Mitchell},
      year={2025},
      eprint={2502.13304},
      archivePrefix={arXiv},
      primaryClass={physics.bio-ph},
      url={https://arxiv.org/abs/2502.13304}, 
}

@misc{huang2024levyscorefunctionscorebased,
      title={L\'{e}vy Score Function and Score-Based Particle Algorithm for Nonlinear L\'{e}vy--Fokker--Planck Equations}, 
      author={Yuanfei Huang and Chengyu Liu and Xiang Zhou},
      year={2024},
      eprint={2412.19520},
      archivePrefix={arXiv},
      primaryClass={math.NA},
      url={https://arxiv.org/abs/2412.19520}, 
}

@article{aerdker2025superdiffusion,
  title={Superdiffusion of energetic particles at shocks: A L{\'e}vy flight model for acceleration},
  author={Aerdker, Sophie and Merten, Lukas and Effenberger, Frederic and Fichtner, Horst and Tjus, Julia Becker},
  journal={Astronomy \& Astrophysics},
  volume={693},
  pages={A15},
  year={2025},
  publisher={EDP Sciences}
}

@article{mantegna1994fast,
  title={Fast, accurate algorithm for numerical simulation of Levy stable stochastic processes},
  author={Mantegna, Rosario Nunzio},
  journal={Physical Review E},
  volume={49},
  number={5},
  pages={4677},
  year={1994},
  publisher={APS},
  note={doi:{\color{blue}\href{http://dx.doi.org/10.1103/physreve.49.4677}{10.1103/physreve.49.4677}}}

}

@article{tang1992pinning,
  title={Pinning by directed percolation},
  author={Tang, Lei-Han and Leschhorn, Heiko},
  journal={Physical Review A},
  volume={45},
  number={12},
  pages={R8309},
  year={1992},
  publisher={APS},
  note={doi:{\color{blue}\href{http://dx.doi.org/10.1103/PhysRevA.45.R8309}{10.1103/PhysRevA.45.R8309}}}
}

@article{harris1974contact,
  title={Contact interactions on a lattice},
  author={Harris, Theodore E},
  journal={The Annals of Probability},
  volume={2},
  number={6},
  pages={969--988},
  year={1974},
  publisher={Institute of Mathematical Statistics},
  note={doi:{\color{blue}\href{http://dx.doi.org/10.1214/aop/1176996493}{10.1214/aop/1176996493}}}

}

@article{pomeau1986front,
  title={Front motion, metastability and subcritical bifurcations in hydrodynamics},
  author={Pomeau, Yves},
  journal={Physica D: Nonlinear Phenomena},
  volume={23},
  number={1-3},
  pages={3--11},
  year={1986},
  publisher={Elsevier},
  note={doi:{\color{blue}\href{http://dx.doi.org/10.1016/0167-2789(86)90104-1}{10.1016/0167-2789(86)90104-1}}}
}

@article{2005Low,
  title={Low-density series expansions for directed percolation IV. Temporal disorder},
  author={ Jensen, Iwan },
  journal={Journal of Physics A General Physics},
  volume={38},
  number={7},
  year={2005},
  note={doi:{\color{blue}\href{http://dx.doi.org/10.1088/0305-4470/38/7/003}{10.1088/0305-4470/38/7/003}}}
}

@article{1998A,
  title={A model for anomalous directed percolation},
  author={Hinrichsen, Haye and Howard, Martin},
  journal={The European Physical Journal B-Condensed Matter and Complex Systems},
  volume={7},
  number={4},
  pages={635--643},
  year={1999},
  publisher={Springer},
  note={doi:{\color{blue}\href{http://dx.doi.org/10.1007/s100510050656}{10.1007/s100510050656}}}
}

@article{hinrichsen2007non,
  title={Non-equilibrium phase transitions with long-range interactions},
  author={Hinrichsen, Haye},
  journal={Journal of Statistical Mechanics: Theory and Experiment},
  volume={2007},
  number={07},
  pages={P07006},
  year={2007},
  publisher={IOP Publishing},
  note={doi:{\color{blue}\href{http://dx.doi.org/10.1088/1742-5468/2007/07/P07006}{10.1088/1742-5468/2007/07/P07006}}}

}

@inproceedings{potts1952some,
  title={Some generalized order-disorder transformations},
  author={Potts, Renfrey Burnard},
  booktitle={Mathematical proceedings of the cambridge philosophical society},
  volume={48},
  number={1},
  pages={106--109},
  year={1952},
  organization={Cambridge University Press},
  note={doi:{\color{blue}\href{http://dx.doi.org/10.1017/S0305004100027419}{10.1017/S0305004100027419}}}

}

@book{tauber2014critical,
  title={Critical dynamics: a field theory approach to equilibrium and non-equilibrium scaling behavior},
  author={T{\"a}uber, Uwe C},
  year={2014},
  publisher={Cambridge University Press}
}

@article{vojta2004broadening,
  title={Broadening of a nonequilibrium phase transition by extended structural defects},
  author={Vojta, Thomas},
  journal={Physical Review E—Statistical, Nonlinear, and Soft Matter Physics},
  volume={70},
  number={2},
  pages={026108},
  year={2004},
  publisher={APS},
  note={doi:{\color{blue}\href{http://dx.doi.org/10.1103/PhysRevE.70.026108}{10.1103/PhysRevE.70.026108}}}
}

@article{hooyberghs2004absorbing,
  title={Absorbing state phase transitions with quenched disorder},
  author={Hooyberghs, Jef and Igl{\'o}i, Ferenc and Vanderzande, Carlo},
  journal={Physical Review E—Statistical, Nonlinear, and Soft Matter Physics},
  volume={69},
  number={6},
  pages={066140},
  year={2004},
  publisher={APS},
  note={doi:{\color{blue}\href{http://dx.doi.org/10.1103/PhysRevE.69.066140}{10.1103/PhysRevE.69.066140}}}
}

@article{cafiero1998disordered,
  title={Disordered one-dimensional contact process},
  author={Cafiero, Raffaele and Gabrielli, Andrea and Mu{\~n}oz, Miguel A},
  journal={Physical Review E},
  volume={57},
  number={5},
  pages={5060},
  year={1998},
  publisher={APS},
  note={doi:{\color{blue}\href{http://dx.doi.org/10.1103/PhysRevE.57.5060}{10.1103/PhysRevE.57.5060}}}
}

@book{1989Quantum,
  title={Quantum field theory and critical phenomena},
  author={Zinn-Justin, Jean},
  volume={171},
  year={2021},
  publisher={Oxford university press}
}

@article{hinrichsen1999flowing,
  title={Flowing sand: A physical realization of directed percolation},
  author={Hinrichsen, Haye and Jim{\'e}nez-Dalmaroni, Andrea and Rozov, Yadin and Domany, Eytan},
  journal={Physical Review Letters},
  volume={83},
  number={24},
  pages={4999},
  year={1999},
  publisher={APS},
  note={doi:{\color{blue}\href{http://dx.doi.org/10.1103/PhysRevLett.83.4999}{10.1103/PhysRevLett.83.4999}}}
}

@article{hinrichsen2000flowing,
  title={Flowing sand—a possible physical realization of Directed Percolation},
  author={Hinrichsen, Haye and Jim{\'e}nez-Dalmaroni, Andrea and Rozov, Yadin and Domany, Eytan},
  journal={Journal of Statistical Physics},
  volume={98},
  pages={1149--1168},
  year={2000},
  publisher={Springer},
  note={doi:{\color{blue}\href{http://dx.doi.org/10.1023/A:1018667712578}{10.1023/A:1018667712578}}}
}

@article{takeuchi2007directed,
  title={Directed percolation criticality in turbulent liquid crystals},
  author={Takeuchi, Kazumasa A and Kuroda, Masafumi and Chat{\'e}, Hugues and Sano, Masaki},
  journal={Physical review letters},
  volume={99},
  number={23},
  pages={234503},
  year={2007},
  publisher={APS},
  note={doi:{\color{blue}\href{http://dx.doi.org/10.1103/PhysRevLett.99.234503}{10.1103/PhysRevLett.99.234503}}}
}

@article{vojta2005critical,
  title={Critical behavior and Griffiths effects in the disordered contact process},
  author={Vojta, Thomas and Dickison, Mark},
  journal={Physical Review E—Statistical, Nonlinear, and Soft Matter Physics},
  volume={72},
  number={3},
  pages={036126},
  year={2005},
  publisher={APS},
  note={doi:{\color{blue}\href{http://dx.doi.org/10.1103/PhysRevE.72.036126}{10.1103/PhysRevE.72.036126}}}
}

@article{gonzaga2019quenched,
  title={Quenched disorder in the contact process on bipartite sublattices},
  author={Gonzaga, MN and Fiore, Carlos Eduardo and de Oliveira, MM},
  journal={Physical Review E},
  volume={99},
  number={4},
  pages={042146},
  year={2019},
  publisher={APS},
 note={doi:{\color{blue}\href{https://doi.org/10.1103/PhysRevE.99.042146}{10.1103/PhysRevE.99.042146}}}


}

@article{dickman2009contact,
  title={A contact process with mobile disorder},
  author={Dickman, Ronald},
  journal={Journal of Statistical Mechanics: Theory and Experiment},
  volume={2009},
  number={08},
  pages={P08016},
  year={2009},
  publisher={IOP Publishing},
  note={doi:{\color{blue}\href{https://doi.org/10.1088/1742-5468/2009/08/P08016}{10.1088/1742-5468/2009/08/P08016}}}



}

@article{fallert2009scaling,
  title={Scaling behavior of the disordered contact process},
  author={Fallert, SV and Taraskin, SN},
  journal={Physical Review E—Statistical, Nonlinear, and Soft Matter Physics},
  volume={79},
  number={4},
  pages={042105},
  year={2009},
  publisher={APS},
  note={doi:{\color{blue}\href{https://doi.org/10.48550/arXiv.0809.0442}{10.48550/arXiv.0809.0442}}}



}

@article{neugebauer2006contact,
  title={Contact process in heterogeneous and weakly disordered systems},
  author={Neugebauer, Chris J and Fallert, SV and Taraskin, SN},
  journal={Physical Review E—Statistical, Nonlinear, and Soft Matter Physics},
  volume={74},
  number={4},
  pages={040101},
  year={2006},
  publisher={APS},
  note={doi:{\color{blue}\href{https://doi.org/10.1103/PhysRevE.74.040101}{10.1103/PhysRevE.74.040101}}}
}

@article{dong2021optimal,
  title={Optimal resilience of modular interacting networks},
  author={Dong, Gaogao and Wang, Fan and Shekhtman, Louis M and Danziger, Michael M and Fan, Jingfang and Du, Ruijin and Liu, Jianguo and Tian, Lixin and Stanley, H Eugene and Havlin, Shlomo},
  journal={Proceedings of the National academy of sciences},
  volume={118},
  number={22},
  pages={e1922831118},
  year={2021},
  publisher={National Academy of Sciences},
  note={doi:{\color{blue}\href{https://doi.org/10.1073/pnas.1922831118}{10.1073/pnas.1922831118}}}
}

@article{liu2021efficient,
  title={Efficient network immunization under limited knowledge},
  author={Liu, Yangyang and Sanhedrai, Hillel and Dong, GaoGao and Shekhtman, Louis M and Wang, Fan and Buldyrev, Sergey V and Havlin, Shlomo},
  journal={National Science Review},
  volume={8},
  number={1},
  pages={nwaa229},
  year={2021},
  publisher={Oxford University Press},
  note={doi:{\color{blue}\href{https://doi.org/10.1093/nsr/nwaa229}{10.1093/nsr/nwaa229}}}

}

@article{qing2024time,
  title={Time persistence of climate and carbon flux networks},
  author={Qing, Ting and Wang, Fan and Li, Qiuyue and Dong, Gaogao and Tian, Lixin and Havlin, Shlomo},
  journal={Communications Physics},
  volume={7},
  number={1},
  pages={372},
  year={2024},
  publisher={Nature Publishing Group UK London},
  note={doi:{\color{blue}\href{https://doi.org/10.1038/s42005-024-01862-9}{10.1038/s42005-024-01862-9}}}
}

\end{document}